\documentclass[notitlepage,a4paper,aps,prd,onecolumn,superscriptaddress,nofootinbib,groupedaddress]{revtex4}
\usepackage{enumerate}
\usepackage{amsmath}
\usepackage{amsfonts}
\usepackage{amssymb}
\usepackage[utf8]{inputenc}
\usepackage[T1]{fontenc}
\usepackage{mathtools}

\usepackage[normalem]{ulem}

\DeclareMathAlphabet{\mathds}{U}{BOONDOX-ds}{m}{n}
\DeclareMathOperator{\arcsinh}{arcsinh}
\usepackage[dvipsnames]{xcolor}

\usepackage[colorlinks=true]{hyperref}
\usepackage{amsthm}

\theoremstyle{definition}
\newtheorem*{defi}{Definition}
\theoremstyle{plain}
\newtheorem{stmt}{Statement}

\DeclareMathOperator{\Vect}{Vect}
\newcommand{\dd}{\mathrm{d}}
\newcommand{\DD}{\mathrm{D}}
\def\Lw{{\mathcal L}{}}

\def\Lw{{\mathcal L}{}}
\def\Lbol{{\stackrel{~\circ}{\mathcal L}}{}}

\begin{document}
\title{Modified teleparallel theories of gravity in symmetric spacetimes}

\author{Manuel Hohmann}
\email{manuel.hohmann@ut.ee}
\affiliation{Laboratory of Theoretical Physics, Institute of Physics, University of Tartu, W. Ostwaldi 1, 50411 Tartu, Estonia.}

\author{Laur J\"arv}
\email{laur.jarv@ut.ee}
\affiliation{Laboratory of Theoretical Physics, Institute of Physics, University of Tartu, W. Ostwaldi 1, 50411 Tartu, Estonia.}

\author{Martin Kr\v{s}\v{s}\'ak}
\email{martin.krssak@ut.ee}
\affiliation{Center for Gravitation and Cosmology, College of Physical Sciences and Technology, Yangzhou University, Yangzhou 225009, China.}
\affiliation{Laboratory of Theoretical Physics, Institute of Physics, University of Tartu, W. Ostwaldi 1, 50411 Tartu, Estonia.}

\author{Christian Pfeifer}
\email{christian.pfeifer@ut.ee}
\affiliation{Laboratory of Theoretical Physics, Institute of Physics, University of Tartu, W. Ostwaldi 1, 50411 Tartu, Estonia.}

\begin{abstract}
Teleparallel gravity theories employ a tetrad and a Lorentz spin connection as independent variables in their covariant formulation. In order to solve their field equations, it is helpful to search for solutions which exhibit certain amounts of symmetry, such as spherical or cosmological symmetry. In this article we present how to apply the notion of spacetime symmetries known from Cartan geometry to teleparallel geometries. We explicitly derive the most general tetrads and spin connections which are compatible with axial, spherical, cosmological and maximal symmetry. For homogeneous and isotropic spacetime symmetry we find that the tetrads and spin connection found by the symmetry constraints are universal solutions to the anti-symmetric part of the field equations of any teleparallel theory of gravity. In other words, for cosmological symmetry we find what has become known as ``good tetrads'' in the context of \(f(T)\) gravity.
\end{abstract}

\maketitle

\section{Introduction}\label{sec:intro}
In the recent decade we have witnessed an increased interest in various extensions of general relativity as a tool to understand the accelerated expansion of the Universe \cite{DeFelice:2010aj,Capozziello:2011et,Nojiri:2017ncd}. One particular class of these models are the so-called \textit{modified teleparallel gravity theories}, where first general relativity was reformulated using the teleparallel geometry to obtain \textit{teleparallel equivalent of general relativity} (TEGR) \cite{Einstein1928,Sauer:2004hj,Moller1961,Hayashi_Nakano1967,Cho1976,Hayashi:1977jd,AP,Krssak:2018ywd}, and then various modified models have been constructed. The teleparallel approach to modified gravity is particularly attractive since it allows us to construct a number of models with second order field equations and hence naturally avoid the problem with higher order field equations faced in other modified theories of gravity. The most popular and the best studied model being the $f(T)$ gravity model, built in analogy with $f(R)$ gravity, where the Lagrangian is considered to be an arbitrary function of the so-called torsion scalar \cite{Ferraro:2006jd,Bengochea:2008gz,Ferraro:2008ey,Linder:2010py,Cai:2015emx}.

A well-known problem of the original formulation of TEGR, in which a zero spin connection is assumed and the tetrad is the only dynamical field, is that the Lagrangian is invariant only under global Lorentz transformations, while local Lorentz transformations introduce a boundary term~\cite{Cho1976}. In the case of modified teleparallel gravity models, such as $f(T)$ gravity, this newly appearing term is not a boundary term, which leads to a violation of the local Lorentz symmetry also at the level of the field equations. As a consequence, considering different tetrads which are related by a local Lorentz transformation, one finds that some of them may solve the field equations, while others may not~\cite{Li:2010cg,Sotiriou:2010mv}. Those tetrads that solved the field equations in a non-trivial way and lead to some interesting new dynamics were named \textit{good tetrads}, and those that could solve the field equations only in the limit of general relativity were named \textit{bad tetrads} \cite{Ferraro:2011us,Tamanini:2012hg}. This introduced a challenge to find these good tetrads in practical situations and also to understand them on the fundamental level.

It was then shown that it is possible to formulate $f(T)$ gravity and other teleparallel models in a fully Lorentz covariant way \cite{Krssak:2015oua,Hohmann:2018rwf,Krssak:2018ywd}, providing that we use the covariant version of TEGR as the base for modifications, where both a tetrad and a teleparallel spin connection are used as fundamental dynamical field variables \cite{AP,Obukhov:2002tm,Obukhov:2006sk,Lucas:2009nq,Krssak_Pereira2015,Krssak:2015lba,Krssak:2017nlv}. While in the covariant TEGR it was argued that the spin connection can be determined only by additional physical requirements, such as the finiteness of the physical action and conserved charges \cite{Krssak:2015lba,Krssak:2017nlv}, in the case of $f(T)$ gravity it was shown that the dynamics of the spin connection are determined by its own field equations, which, moreover, coincide with the antisymmetric part of the field equations for the tetrad \cite{Golovnev:2017dox}. Recently, it was shown that the property extends to all modified teleparallel theories with second order field equations \cite{Hohmann:2017duq}.

Nevertheless, it remained an open problem of how to determine both, the tetrad and the spin connection in practical situations since (even the antisymmetric part of) the field equations are typically too complicated to be solved in general. It was known that it is possible to ``guess'' some ansatzes for the spin connection in the case of spherical symmetry and the spatially flat Friedmann-Lemaitre-Robertson-Walker universe, which turned out to be equivalent to guessing the good tetrads in the non-covariant formulation, but it remained unclear how to generalize this procedure \cite{Krssak:2015oua}.

In this paper, we present a notion of symmetry for teleparallel geometries on which teleparallel theories of gravity are based. We obtain a precise notion of symmetry for the fundamental fields, the tetrad and the spin connection, by mapping the teleparallel geometry to a Cartan geometry and adopting the known symmetry principles from there~\cite{Hohmann:2015pva}. Furthermore we show that the torsion, and thus the resulting teleparallel field equations, inherit the symmetry imposed on the fundamental fields.

We demonstrate the use of our method in the case of axially, spherically symmetric spacetimes as well as homogeneous and isotropic and maximally symmetric spacetimes. In the homogeneous and isotropic case we will find that it is possible to find a tetrad and a spin connection that solve the antisymmetric part of the tetrad field equations for all teleparallel models with second order field equations, derived purely on the basis of the symmetry considerations. In other words, for cosmology we find a universal good tetrad. We then also examine the case of maximally symmetric spacetime and demonstrate that such general solutions exist only in the case of Minkowski spacetime. Our method proves that in the case of (anti-)de Sitter spacetime there is no tetrad and flat spin connection that would solve the symmetry conditions. We would like to stress here that this does not exclude the possibility that there are teleparallel geometries which realize the (anti-)de Sitter spacetime metric and solve the field equations in some specific models, e.g. $f(T)$ gravity.

The outline of the paper is as follows. In section~\ref{sec:covtelegrav}, we review teleparallel geometry and introduce the covariant formulation of TEGR and other modified teleparallel gravity models. In section~\ref{sec:symtelegrav} we introduce the notion of symmetry in teleparallel geometries and show how to construct the tetrad and the spin connection leading to the torsion tensor exhibiting the desired symmetry.
In section~\ref{sec:solutions} we demonstrate the use of this method in the case of axially and spherically symmetric spacetimes, and both spatially flat and non-flat FRWL Universes, and we discuss the case of maximally symmetric spacetimes. In particular we find in section \ref{sssec:cosmogen}, that for high-symmetric situations, the tetrads and spin connections which satisfy the symmetry conditions automatically solve the antisymmetric, resp. the spin connection, field equation for all the models mentioned in section \ref{sec:covtelegrav}. Finally in section~\ref{sec:conclusion} we conclude and summarize the main results.

\section{Covariant formulation of teleparallel gravity}\label{sec:covtelegrav}
This section introduces and recalls the basic notions on teleparallel geometry in subsection \ref{ssec:telegeom} and teleparallel theories of gravity in subsection \ref{ssec:tpgrav}, which we need throughout this article.

\subsection{Teleparallel geometry}\label{ssec:telegeom}
We start our discussion with a brief review of the mathematical notions we use in order to describe teleparallel geometry. In this article we use the covariant formulation~\cite{AP,Krssak:2018ywd}, where the geometry given on a spacetime manifold \(M\) is defined in terms of a tetrad \(\boldsymbol{\theta} \in \Omega^1(M, \mathfrak{z})\) and a spin connection \(\boldsymbol{\omega} \in \Omega^1(M, \mathfrak{h})\). Both are $1$-forms on \(M\), the former with values in Minkowski space \(\mathfrak{z} = \mathbb{R}^{1,3}\) equipped with a bilinear form \(\eta\) of signature \((-+++)\), the latter with values in the Lorentz algebra \(\mathfrak{h} = \mathfrak{so}(1,3)\). We denote the teleparallel geometry by the triple \((M, \boldsymbol{\theta}, \boldsymbol{\omega})\).

By introducing coordinates \((x^{\mu})\) on \(M\) and suitable bases of \(\mathfrak{z}\) and \(\mathfrak{h}\), we can write $\boldsymbol{\theta} = \{\theta^a\}_{a=0}^3$ as \(\theta^a = \theta^a{}_{\mu}\dd x^\mu\) and $\boldsymbol{\omega} = \{\omega^a{}_b\}_{a,b=0}^3$ as \(\omega^a{}_b = \omega^a{}_{b\mu} \dd x^\mu\). Here we denoted Lorentz indices with Latin letters \(a, b, \ldots\), while Greek indices \(\mu, \nu, \ldots\) denote tangent space indices of \(M\). In order to be teleparallel, the spin connection must be flat, i.e., it must satisfy the vanishing curvature condition
\begin{equation}\label{eq:curvfree}
	R^a{}_{b\mu\nu} = \partial_{\mu}\omega^a{}_{b\nu} - \partial_{\nu}\omega^a{}_{b\mu} + \omega^a{}_{c\mu}\omega^c{}_{b\nu} - \omega^a{}_{c\nu}\omega^c{}_{b\mu} = 0\,.
\end{equation}

From the tetrad and the connection one defines a metric
\begin{equation}\label{eq:metric}
	g_{\mu\nu} = \eta_{ab}\theta^a{}_{\mu}\theta^b{}_{\nu}
\end{equation}
and an affine connection \(\nabla\) with connection coefficients
\begin{equation}\label{eq:affconn}
	\Gamma^{\mu}{}_{\nu\rho} = e_a{}^{\mu}\left(\partial_{\rho}\theta^a{}_{\nu} + \omega^a{}_{b\rho}\theta^b{}_{\nu}\right)\,,
\end{equation}
where \(e_a{}^{\mu}\) is the inverse tetrad satisfying \(\theta^a{}_{\mu}e_b{}^{\mu} = \delta^a_b\) and \(\theta^a{}_{\mu}e_a{}^{\nu} = \delta_{\mu}^{\nu}\). For tensor components the tetrad and its inverse can be used to transform Lorentz indices into spacetime indices and vice versa.

Being obtained from the flat spin connection, the affine connection also has vanishing curvature,
\begin{equation}\label{eq:affcurvfree}
	R^{\rho}{}_{\sigma\mu\nu} = \partial_{\mu}\Gamma^{\rho}{}_{\sigma\nu} - \partial_{\nu}\Gamma^{\rho}{}_{\sigma\mu} + \Gamma^{\rho}{}_{\tau\mu}\Gamma^{\tau}{}_{\sigma\nu} - \Gamma^{\rho}{}_{\tau\nu}\Gamma^{\tau}{}_{\sigma\mu} = 0\,,
\end{equation}
but in general non-vanishing torsion
\begin{equation}\label{eq:torsion}
	T^{\rho}{}_{\mu\nu} = \Gamma^{\rho}{}_{\nu\mu} - \Gamma^{\rho}{}_{\mu\nu}\,.
\end{equation}
As a consequence of \eqref{eq:affconn}, we find that the total covariant derivative, which acts on Lorentz and spacetime indices,
\begin{equation}
	\DD_{\mu}\theta^a{}_{\nu} = \partial_{\mu}\theta^a{}_{\nu} + \omega^a{}_{b\mu}\theta^b{}_{\nu} - \Gamma^{\rho}{}_{\nu\mu}\theta^a{}_{\rho}
\end{equation}
of the tetrad vanishes.

We now consider a local Lorentz transformation \(\Lambda: M \to \mathrm{SO}(1,3)\), under which the tetrad and the spin connection transform as
\begin{equation}
	\theta^a{}_{\mu} \mapsto \Lambda^a{}_b\theta^b{}_{\mu}\,, \quad \omega^a{}_{b\mu} \mapsto \Lambda^a{}_c\Lambda_b{}^d\omega^c{}_{d\mu} + \Lambda^a{}_c\partial_{\mu}\Lambda_b{}^c\,,
\end{equation}
with the inverse Lorentz transform given by \(\Lambda_a{}^b = (\Lambda^{-1})^b{}_a\). One easily checks that the condition of vanishing curvature is unaffected by this transformation, since
\begin{equation}
	R^a{}_{b\mu\nu} \mapsto \Lambda^a{}_c\Lambda_b{}^dR^c{}_{d\mu\nu}\,.
\end{equation}
Further, the metric \(g_{\mu\nu}\) and connection coefficients \(\Gamma^{\mu}{}_{\nu\rho}\) (and hence also the curvature \(R^{\rho}{}_{\sigma\mu\nu}\) and torsion \(T^{\rho}{}_{\mu\nu}\)) are invariant under this transformation. We finally remark that the most general metric-compatible spin connection with vanishing curvature is itself of the ``purely inertial'' form
\begin{equation}\label{telespin}
	\omega^a{}_{b\mu} = \Lambda^a{}_c\partial_{\mu}\Lambda_b{}^c
\end{equation}
for some local Lorentz transformation, and can hence always be transformed into a zero spin connection by applying the inverse Lorentz transformation.

\subsection{Teleparallel theories of gravity}\label{ssec:tpgrav}
Using the teleparallel geometry we can construct various theories of gravity. In particular, we can consider the so-called \textit{teleparallel equivalent of general relativity} (TEGR), often referred simply as just \textit{teleparallel gravity}, by considering a Lagrangian \cite{AP}
\begin{equation}
\Lw_\text{TG}(\theta^a_{\ \mu},\omega^a_{\ b\mu})=\frac{\theta}{2 \kappa} T, \label{lagtot}
\end{equation}
where $\theta=\det \theta^a_{\ \mu}$, $\kappa=8\pi G$ is the gravitational constant (in $c=1$ units), and $T$ is a quadratic combination of torsion tensors
\begin{equation}
	T=T(\theta^a_{\ \mu},\omega^a_{\ b\mu})= \frac{1}{4} \; T^\rho{}_{\mu\nu} \, T_\rho{}^{\mu \nu} +
	\frac{1}{2}  \; T^\rho{}_{\mu\nu} \, T^{\nu \mu}{}_\rho -
	\, T^\rho{}_{\mu\rho} \, T^{\nu \mu}{}_\nu,
	\label{tscalar}
\end{equation}
known as the torsion scalar. It is possible to rewrite the torsion scalar as
\begin{equation}
	T= \frac{1}{2} T^a_{\ \mu\nu}S_a^{\ \mu\nu},
\end{equation}
where we have defined the superpotential $S_a^{\ \rho\sigma}$ as
\begin{equation}
	S_a^{\ \mu\nu}=\frac{1}{2}
	\left(
	T^{\nu \mu}_{\ \ \ a}
	+T^{\ \mu\nu}_{a}
	-T^{\mu\nu}_{\ \ \ a}
	\right)
	-h_a^{\ \nu}T^{\sigma \mu}_{\ \ \sigma}
	+h_a^{\ \mu}T^{\sigma \nu}_{\ \ \sigma}.
	\label{sup}
\end{equation}
We then consider  the total Lagrangian
\begin{equation}\label{actotal}
	\Lw (\theta^a_{\ \mu},\omega^a_{\ b\mu},\Phi^I)= \Lw_\text{TG} (\theta^a_{\ \mu},\omega^a_{\ b\mu}) + \Lw_\text{M}(\theta^a_{\ \mu},\Phi^I),
\end{equation}
where $\Lw_\text{M}(\theta^a_{\ \mu},\Phi^I)$ is the matter Lagrangian  constructed through the minimal coupling principle and $\Phi^I$ denotes various matter fields. The variation of the total Lagrangian \eqref{actotal} with respect to the tetrad yields the field equations
\begin{equation}
	E_a^{\ \mu} = \kappa \,  {\Theta}_{a}{}^{\mu},
	\label{fe}
\end{equation}
where on the left-hand side we have defined the Euler-Lagrange expression
\begin{equation}
	E_a^{\ \mu} \equiv \frac{\kappa}{\theta} \frac{\delta \Lw_\text{TG}}{\delta \theta^a_{\ \mu}} =\frac{1}{\theta}\, \partial_\sigma\Big(\theta S_a{}^{\mu \sigma}\Big) -
	\theta_a{}^{\lambda}S_c{}^{\nu\mu}T^c{}_{\nu\lambda} +\frac{e_a^{\ \mu}}{2} T -  \omega^c{}_{a\sigma} S_c{}^{\mu\sigma},
\end{equation}
and on the right-hand side is the  matter energy-momentum tensor
\begin{equation}\label{matemten}
	{\Theta}_{a}{}^{\mu} = - \frac{1}{\theta} \frac{\delta {\mathcal L}_\text{M}}{\delta \theta^a_{\ \mu}},
\end{equation}
which is symmetric as a consequence of the local Lorentz invariance of the action~\cite{AP,Obukhov:2006sk}.

It can be shown \cite{AP} that the teleparallel Lagrangian \eqref{lagtot} is (up to a surface term) equivalent to the Einstein-Hilbert Lagrangian $\Lbol_{EH}$, which is given by the Ricci-Scalar of the Levi-Civita connection of the metric, understood as function of the tetrads, i.e.
\begin{equation}\label{lagequiv}
	\Lw_{TG} \equiv \Lbol_{EH}  - \frac{1}{\kappa}\partial_\mu \Big(\theta T^{\rho\mu}{}_\rho\Big),
\end{equation}
from where follows that the field equations \eqref{fe} are equivalent to Einstein field equations of general relativity.

The variation of the Lagrangian with respect to the flat spin connection vanishes identically \cite{Krssak2017,Golovnev:2017dox,Krssak:2017nlv}, and hence the field equations for the spin connection are identically satisfied. The spin connection in this particular  model equivalent to general relativity can be determined by the requirement of finiteness of the action and conserved charges \cite{Krssak_Pereira2015}.

An intriguing property of the Lagrangian \eqref{lagtot} is that it contains only the first derivatives of field variables, what allows us to construct various modified gravity models with second order field equations and hence avoid problems with higher order field equations faced in modified gravity models built on the standard general relativity. The most popular among these models is the so-called $f(T)$ gravity model \cite{Ferraro:2006jd,Bengochea:2008gz,Ferraro:2008ey,Linder:2010py}, where the Lagrangian is taken to be an arbitrary function of the torsion scalar \eqref{tscalar}, i.e.
\begin{equation}\label{ftaction}
	\Lw_{f}=\frac{\theta}{2\kappa} f(T).
\end{equation}
The field equations for the tetrad can be written as \eqref{fe} with the Euler-Lagrange expression given by \cite{Krssak:2015oua}
\begin{equation}\label{ftfe}
	E_a^{\ \mu}\equiv \frac{\kappa}{\theta} \frac{\delta \Lw_f}{\delta \theta^a_{\ \mu}} = \frac{1}{\theta}f_T \partial_{\nu}\left( \theta S_a^{\ \mu\nu} \right)
	+
	f_{TT} S_a^{\ \mu\nu} \partial_{\nu} T
	-
	f_T T^b_{\ \nu a }S_b^{\ \nu\mu}
	+
	f_T \omega^b_{\ a\nu}S_b^{\ \nu\mu}
	+
	\frac{1}{2}f e_a^{\ \mu},
\end{equation}
where $f_T$ and $f_{TT}$ denotes first and second order derivatives of the $f$-function with respect to the torsion scalar.

Unlike the case of TEGR, the field equations of $f(T)$ gravity \eqref{ftfe} are generally not symmetric. Defining the fully Lorentz-indexed Euler-Lagrange expression as $E_{ab}= \eta_{bc} \theta^c_{\ \mu}E_a^{\ \mu}$, we can write the antisymmetric part of the field equations as
\begin{equation}\label{feas}
	E_{[ab]}=\theta\, f_{TT}S_{[ab]}{}^\nu \partial_\nu T.
\end{equation}

Another important difference with the TEGR case is that the variation with respect to the flat spin connection is non-trivial, and hence the dynamics of the spin connection is determined by its own field equations. We can write the variation with respect to the flat spin connection as
\begin{equation}\label{ftvar}
	\delta_\omega  \Lw_f= \frac{\theta}{2 \kappa} f_T S_{ab}{}^{\mu} {\delta \omega^{a b}{}_{\mu}}.
\end{equation}
In order to preserve the flatness of the spin connection and anti-symmetry in its Lorentz indices, it is necessary to consider  the changes of the spin connection under infinitesimal  local Lorentz transformations only
\begin{equation}
	\Lambda^a_{\ b}=\delta^a_{\ b}+\epsilon^a_{\ b}, \qquad \qquad \epsilon_{ab}=-\epsilon_{ba},
\end{equation}
from where we find that
\begin{equation}\label{convarr}
	\delta \omega^{ab}{}_{\mu}=-\DD_\mu \epsilon^{ab}=-\partial_\mu \epsilon^{ab} - \omega^a{}_{c\mu}\epsilon^{cb} - \omega^b{}_{c\mu} \epsilon^{ac}\,.
\end{equation}

Using \eqref{convarr} and integrating by parts, we find that the variation with respect to the spin connection \eqref{ftvar} lead to the same result as the antisymmetric field equations for the tetrad \cite{Golovnev:2017dox,Hohmann:2017duq}. This implies that if the field equations for the spin connection are satisfied, the field equations for the tetrad are guaranteed to be symmetric. As we will show in this paper, in the situations with enough symmetry, it is possible to solve the spin connection field equations independently of the symmetric field equations for the tetrad.

The case of $f(T)$ gravity is the most popular and best studied teleparallel gravity model, but it is possible to construct a  number of other interesting modified gravity theories within teleparallel framework. As it has been shown recently~\cite{Hohmann:2017duq}, all teleparallel gravity models with second order field equations have this property that the field equations for the spin connection coincide with the antisymmetric part of the tetrad field equations. We  outline here few of the most popular models to which this property applies. For further details, see the recent review \cite{Krssak:2018ywd}.

The earliest modified teleparallel model is known as \textit{new general relativity} \cite{Hayashi_Shirafuji1979,Hayashi_Shirafuji1982}, where arbitrary coefficients in front of the terms in the torsion scalar \eqref{tscalar} are considered. Following the original approach \cite{Hayashi_Shirafuji1979,Hayashi_Shirafuji1982}, we can  decompose the torsion tensor into irreducible parts with respect to the Lorentz group as
\begin{align}
	T_{\lambda\mu\nu} = \frac{2}{3}(t_{\lambda\mu\nu}-t_{\lambda\nu\mu}) +
	\frac{1}{3}(g_{\lambda\mu}v_{\nu}-g_{\lambda\nu}v_{\mu}) +
	\epsilon_{\lambda\mu\nu\rho}a^{\rho}\,,
\end{align}
where
\begin{align}
	v_{\mu} = T^{\lambda}{}_{\lambda\mu}\,,\qquad
	a_{\mu} = \frac{1}{6}\epsilon_{\mu\nu\sigma\rho}T^{\nu\sigma\rho}\,, \qquad
	t_{\lambda\mu\nu} = \frac{1}{2}(T_{\lambda\mu\nu}+T_{\mu\lambda\nu}) +
	\frac{1}{6}(g_{\nu\lambda}v_{\mu}+g_{\nu\mu}v_{\lambda})-\frac{1}{3}g_{\lambda\mu}v_{\nu}\,,
\end{align}
are known as the vector, axial, and purely tensorial, torsions, respectively.

Constructing the following three parity preserving quadratic invariants
\begin{equation}
	T_{\rm vec} = v_{\mu}v^{\mu},\qquad
	T_{\rm ax} = a_{\mu}a^{\mu}, \qquad
	T_{\rm ten} = t_{\lambda\mu\nu}t^{\lambda\mu\nu},
	\label{qinv}
\end{equation}
the Lagrangian of new general relativity is then written as
\begin{equation}
	\Lw_{\rm NGR} =
	-\frac{\theta}{2\kappa}\Big(a_{0} + a_{1} T_{\rm ax} + a_{2} T_{\rm ten} + a_{3} T_{\rm vec} \Big).
	\label{ngraction}
\end{equation}

Recently \cite{Bahamonde:2017wwk},  it was shown that it is possible to consider a general model known as $f(T_\text{ax},T_\text{ten},T_\text{vec})$ gravity, where we consider a Lagrangian given by an arbitrary function of the quadratic invariants \eqref{qinv}
\begin{align}
	\Lw_{ATV} = \frac{\theta}{2\kappa} \,
	f(T_{\rm ax},T_{\rm ten},T_{\rm vec}).
	\label{action3}
\end{align}

It is also possible to modify teleparallel gravity by introducing a non-minimal coupling with the scalar field $\phi$ \cite{Geng:2011aj,Hohmann:2018rwf}, which can be further generalized to the so-called $f(T,X,Y,\phi)$ model \cite{Hohmann:2018dqh},  where $X$ is the kinetic term for  scalar field, and $Y$ is the term representing the coupling between the torsion and gradient of the scalar field.

\section{Symmetries in teleparallel gravity}\label{sec:symtelegrav}
We now come to the general notion of spacetime symmetries in the language of teleparallel geometry, which we will investigate in several steps. We start with a definition of symmetry under a finite group action in section~\ref{ssec:finisym}. The infinitesimal version of this notion will be developed in section~\ref{ssec:infisym}. We then discuss how this notion of symmetry behaves under local Lorentz transformations in section~\ref{ssec:loclor}, and make use of these properties in order to simplify the symmetry conditions. Finally, in section~\ref{ssec:symfeq} we discuss how this notion of symmetry leads to a simplification of the field equations of teleparallel gravity theories. Observe that the results in the sections \ref{ssec:finisym} and \ref{ssec:infisym} do not yet use the teleparallel condition of the connection, but hold for general metric-compatible spin connections. Only in section \ref{ssec:loclor} the curvature free condition will be used to establish results in the Weitzenböck gauge.

\subsection{Symmetries under group actions}\label{ssec:finisym}

There are different possible ways to define spacetime symmetries. For our purposes it will be most convenient to consider the invariance of selected geometric objects on a spacetime manifold \(M\) under the action \(\varphi: G \times M \to M\) of a Lie group \(G\) on \(M\). The particular notion of symmetry we use here is motivated by a previous study of spacetime symmetries using Cartan geometry~\cite{Hohmann:2015pva}, and is also valid in the more general case when the spin connection \(\omega\) is not flat\footnote{Let us mention that recently it has been discussed whether the Cartan geometry in teleparallel geometry should be used at the more fundamental level in Refs.~\cite{Fontanini:2018krt} and~\cite{Pereira:2019woq}. In our paper, we use Cartan geometry only as a mathematical tool since Cartan geometry provides a possible interpretation of the tetrad and spin connection used in the teleparallel geometry.}. The teleparallel fields tetrad and spin connection, introduced in section~\ref{ssec:telegeom}, define via equations \eqref{eq:metric} and \eqref{eq:affconn} a metric and an affine connection. They in turn define what is called an orthogonal Cartan geometry, for which a precise notion of symmetry exists. The orthogonal Cartan geometry is invariant under a group action on \(M\) if and only the metric \(g_{\mu\nu}\) and the affine connection \(\nabla\) are invariant under this action. To clarify this notion, let \(u \in G\) and denote the induced diffeomorphism by \(\varphi_u: M \to M\). Further denoting the image \(\varphi_u(x)\) by \(x'\), one finds the usual formulas
\begin{equation}
(\varphi_u^*g)_{\mu\nu}(x) = g_{\rho\sigma}(x')\frac{\partial x'^{\rho}}{\partial x^{\mu}}\frac{\partial x'^{\sigma}}{\partial x^{\nu}}
\end{equation}
and
\begin{equation}
(\varphi_u^*\Gamma)^\mu{}_{\nu\rho}(x) = \Gamma^{\tau}{}_{\omega\sigma}(x')\frac{\partial x^{\mu}}{\partial x'^{\tau}}\frac{\partial x'^{\omega}}{\partial x^{\nu}}\frac{\partial x'^{\sigma}}{\partial x^{\rho}} + \frac{\partial x^{\mu}}{\partial x'^{\sigma}}\frac{\partial^2x'^{\sigma}}{\partial x^{\nu}\partial x^{\rho}}
\end{equation}
for the pullbacks of the metric and the connection coefficients along the diffeomorphism \(\varphi_u\). Following the treatment in~\cite{Hohmann:2015pva} we introduce the following notion of symmetry.

\begin{defi}
A \emph{symmetry} of a teleparallel geometry \((M, \boldsymbol{\theta},\boldsymbol{\omega})\) is a group action \(\varphi: G \times M \to M\) of a Lie group \(G\) such that the induced metric~\eqref{eq:metric} and affine connection~\eqref{eq:affconn} are invariant, i.e., \(\varphi_u^*g = g\) and \(\varphi_u^*\Gamma = \Gamma\) for all \(u \in G\). The teleparallel geometry is then called \emph{symmetric} under the group action \(\varphi\).
\end{defi}

We now express the conditions of symmetry on the metric and affine connection as conditions on the fundamental teleparallel variables tetrad and spin connection. Observe that their pullbacks by \(\varphi_u\) are given by
\begin{equation}
(\varphi_u^*\theta)^a{}_{\mu}(x) = \theta^a{}_{\nu}(x')\frac{\partial x'^{\nu}}{\partial x^{\mu}}\,, \quad (\varphi_u^*\omega)^a{}_{b\mu}(x) = \omega^a{}_{b\nu}(x')\frac{\partial x'^{\nu}}{\partial x^{\mu}}\,,
\end{equation}
as usual for $1$-forms. It follows from their definitions~\eqref{eq:metric} and~\eqref{eq:affconn} that the metric and the affine connection are invariant under the diffeomorphism \(\varphi_u\), i.e, \(\varphi_u^*g = g\) and \(\varphi_u^*\Gamma = \Gamma\), if and only if there exists a corresponding local Lorentz transformation \(\boldsymbol{\tilde{\Lambda}}_u: M \to \mathrm{SO}(1,3)\) such that
\begin{equation}\label{eq:finisymcondgen}
(\varphi_u^*\theta)^a{}_{\mu} = \boldsymbol{\tilde{\Lambda}}_u^a{}_b\theta^b{}_{\mu}\,, \quad
(\varphi_u^*\omega)^a{}_{b\mu} = \boldsymbol{\tilde{\Lambda}}_u^a{}_c\boldsymbol{\tilde{\Lambda}}_{u\,b}{}^d\omega^c{}_{d\mu} + \boldsymbol{\tilde{\Lambda}}_u^a{}_c\partial_{\mu}\boldsymbol{\tilde{\Lambda}}_{u\,b}{}^c\,.
\end{equation}
Hence, the metric and connection are invariant under the action of the group \(G\) if and only if there exists a map \(\boldsymbol{\tilde{\Lambda}}: G \times M \to \mathrm{SO(1,3)}\) such that this condition is satisfied for every \(u \in G\).

From the fact that we are considering a (left) group action follows that any such map \(\boldsymbol{\tilde{\Lambda}}\) which satisfies the condition~\eqref{eq:finisymcondgen} has another property, which is imposed by the group structure of \(G\). For \(u,v \in G\) we find
\begin{equation}
\boldsymbol{\tilde{\Lambda}}_{uv}^a{}_b\theta^b{}_{\mu} = (\varphi_{uv}^*\theta)^a{}_{\mu} = (\varphi_v^*\varphi_u^*\theta)^a{}_{\mu} = \boldsymbol{\tilde{\Lambda}}_v^a{}_b\boldsymbol{\tilde{\Lambda}}_u^b{}_c\theta^c{}_{\mu} \quad \Rightarrow \quad \boldsymbol{\tilde{\Lambda}}_{uv} = \boldsymbol{\tilde{\Lambda}}_v\boldsymbol{\tilde{\Lambda}}_u\,,
\end{equation}
where the last part follows after multiplication with the inverse tetrad. This shows that \(\boldsymbol{\tilde{\Lambda}}\) is a local homomorphism of the opposite group of \(G\) to the Lorentz group \(\mathrm{SO}(1,3)\). The same result can also be derived from the transformation behavior of \(\omega\). This also shows that it is more convenient to consider the inverse \(\boldsymbol{\Lambda} = \bullet^{-1} \circ \boldsymbol{\tilde{\Lambda}}\) instead, which is a local homomorphism of \(G\) to \(\mathrm{SO}(1,3)\), and whose components are given by
\begin{equation}
\boldsymbol{\Lambda}_u^a{}_b = \boldsymbol{\tilde{\Lambda}}_{u^{-1}}^a{}_b = \left(\boldsymbol{\tilde{\Lambda}}_u^{-1}\right)^a{}_b = \boldsymbol{\tilde{\Lambda}}_{u\,b}{}^a\,.
\end{equation}
This leads us to the following statement:

\begin{stmt}
A teleparallel geometry \((M, \boldsymbol{\theta},\boldsymbol{\omega})\) is symmetric under a group action \(\varphi: G \times M \to M\) if and only if there exists a local Lie group homomorphism \(\boldsymbol{\Lambda}: G \times M \to \mathrm{SO}(1,3)\) such that the conditions~\eqref{eq:finisymcondgen} are satisfied for \(\boldsymbol{\tilde{\Lambda}} =  \bullet^{-1} \circ \boldsymbol{\Lambda}\) and for all \(u \in G\).
\end{stmt}

In the following we will only work with \(\boldsymbol{\Lambda}\) instead of \(\boldsymbol{\tilde{\Lambda}}\).

\subsection{Infinitesimal symmetries}\label{ssec:infisym}

For practical calculations it is often convenient to consider infinitesimal symmetries instead of the finite symmetries discussed above. Recall that any element \(\xi \in \mathfrak{g} \cong T_{\mathds{1}}G\) determines a one-parameter subgroup \(\hat{\xi}: \mathbb{R} \to G, t \mapsto \exp(t\xi)\) via the exponential map. The action \(\varphi: G \times M \to M\) assigns to \(\xi\) a one-parameter group \(\varphi_{\hat{\xi}(t)}\) of diffeomorphisms, which determines a trajectory \(\gamma_{\xi}(x,t) = \varphi_{\hat{\xi}(t)}(x)\), such that \(\gamma_{\xi}(x, 0) = x\). The tangent vectors \(\dot{\gamma}_{\xi}(x, 0) = X_{\xi}(x)\) constitute a vector field \(X_{\xi} \in \Vect M\). Note that the assignment \(\xi \mapsto X_{\xi}\) is a Lie algebra homomorphism.

The Lie derivative of a tensor field \(W\) or affine connection on \(M\) with respect to \(X_{\xi}\) is given by
\begin{equation}
\mathcal{L}_{X_{\xi}}W = \left.\frac{\dd}{\dd t}\left(\varphi_{\hat{\xi}(t)}^*W\right)\right|_{t = 0}\,.
\end{equation}
In particular, for the metric and affine connection we have the Killing equation
\begin{equation}\label{eq:metsymcond}
(\mathcal{L}_{X_{\xi}}g)_{\mu\nu} = X_{\xi}^{\rho}\partial_{\rho}g_{\mu\nu} + \partial_{\mu}X_{\xi}^{\rho}g_{\rho\nu} + \partial_{\nu}X_{\xi}^{\rho}g_{\mu\rho}
\end{equation}
and
\begin{equation}\label{eq:affsymcond}
\begin{split}
(\mathcal{L}_{X_{\xi}}\Gamma)^{\mu}{}_{\nu\rho} &= X_{\xi}^{\sigma}\partial_{\sigma}\Gamma^{\mu}{}_{\nu\rho} - \partial_{\sigma}X_{\xi}^{\mu}\Gamma^{\sigma}{}_{\nu\rho} + \partial_{\nu}X_{\xi}^{\sigma}\Gamma^{\mu}{}_{\sigma\rho} + \partial_{\rho}X_{\xi}^{\sigma}\Gamma^{\mu}{}_{\nu\sigma} + \partial_{\nu}\partial_{\rho}X_{\xi}^{\mu}\\
&= \nabla_{\rho}\nabla_{\nu}X_{\xi}^{\mu} - X_{\xi}^{\sigma}R^{\mu}{}_{\nu\rho\sigma} - \nabla_{\rho}(X_{\xi}^{\sigma}T^{\mu}{}_{\nu\sigma})\,,
\end{split}
\end{equation}
where the last line shows that \(\mathcal{L}_{X_{\xi}}\Gamma\) is a tensor field. (Note that the equation given here holds also for non-flat connection \(\nabla\), while in the teleparallel case one has \(R^{\mu}{}_{\nu\rho\sigma} \equiv 0\).) The tetrad and spin connection transform as $1$-forms, hence
\begin{equation}
(\mathcal{L}_{X_{\xi}}\theta)^a{}_{\mu} = X_{\xi}^{\nu}\partial_{\nu}\theta^a{}_{\mu} + \partial_{\mu}X_{\xi}^{\nu}\theta^a{}_{\nu}\,, \quad (\mathcal{L}_{X_{\xi}}\omega)^a{}_{b\mu} = X_{\xi}^{\nu}\partial_{\nu}\omega^a{}_{b\mu} + \partial_{\mu}X_{\xi}^{\nu}\omega^a{}_{b\nu}\,.
\end{equation}

If the metric and connection are invariant under the action of the group \(G\), then the Lie derivatives \(\mathcal{L}_{X_{\xi}}g\) and \(\mathcal{L}_{X_{\xi}}\Gamma\) vanish. This is the case if and only if the Lie derivative of the tetrad and the spin connection satisfy
\begin{equation}\label{eq:infisymcondgen}
(\mathcal{L}_{X_{\xi}}\theta)^a{}_{\mu} = -\boldsymbol{\lambda}_{\xi}^a{}_b\theta^b{}_{\mu}\,, \quad
(\mathcal{L}_{X_{\xi}}\omega)^a{}_{b\mu} = \DD_{\mu}\boldsymbol{\lambda}_{\xi}^a{}_b\,,
\end{equation}
where we used the total derivative
\begin{equation}
\DD_{\mu}\boldsymbol{\lambda}_{\xi}^a{}_b = \partial_{\mu}\boldsymbol{\lambda}_{\xi}^a{}_b + \omega^a{}_{c\mu}\boldsymbol{\lambda}_{\xi}^c{}_b - \omega^c{}_{b\mu}\boldsymbol{\lambda}_{\xi}^a{}_c\,,
\end{equation}
and where \(\boldsymbol{\lambda}: \mathfrak{g} \times M \to \mathfrak{so}(1,3)\) is the local Lie algebra homomorphism defined by
\begin{equation}\label{eq:lochomo}
\boldsymbol{\lambda}_{\xi}(x) = \left.\frac{\dd}{\dd t}\boldsymbol{\Lambda}_{\hat{\xi}(t)}(x)\right|_{t = 0}\,.
\end{equation}

\subsection{Local Lorentz transformations}\label{ssec:loclor}

We now consider a tetrad \(\theta'\) and spin connection \(\omega'\) derived from a tetrad \(\theta\) and spin connection \(\omega\) by a local Lorentz transformation \(\Lambda: M \to \mathrm{SO}(1,3)\) such that
\begin{equation}
\theta'^a{}_{\mu} = \Lambda^a{}_b\theta^b{}_{\mu}\,, \quad \omega'^a{}_{b\mu} = \Lambda^a{}_c\Lambda_b{}^d\omega^c{}_{d\mu} + \Lambda^a{}_c\partial_{\mu}\Lambda_b{}^c\,.
\end{equation}
Recall that a local Lorentz transformation of this form leaves the metric and affine connection invariant, \(g = g'\) and \(\nabla = \nabla'\). Hence, if \(\theta\) and \(\omega\) are symmetric under the action of a group \(G\) on \(M\) with local homomorphism \(\boldsymbol{\Lambda}\), then there exists also a local homomorphism \(\boldsymbol{\Lambda'}\) such that \(\theta'\) and \(\omega'\) are symmetric. A quick calculation shows that this local homomorphism is given by
\begin{equation}
\boldsymbol{\Lambda}'_{u\,b}{}^a(x) = \Lambda^a{}_c(\varphi_u(x))\Lambda_b{}^d(x)\boldsymbol{\Lambda}_{u\,d}{}^c(x)\,.
\end{equation}

For teleparallel connections satisfying \eqref{eq:curvfree}, and thus being of the form \eqref{telespin}, we may use this fact in order to simplify the symmetry conditions~\eqref{eq:finisymcondgen}, or their infinitesimal versions~\eqref{eq:infisymcondgen}. It allows us to work in the Weitzenböck gauge, i.e., to choose \(\Lambda^a{}_b\) such that \(\omega'^a{}_{b\mu} = 0\). In this gauge the symmetry conditions~\eqref{eq:finisymcondgen} read
\begin{equation}\label{eq:finisymcondwb}
(\varphi_u^*\theta')^a{}_{\mu} = \boldsymbol{\Lambda}'_{u\,b}{}^a\theta'^b{}_{\mu}\,, \quad
0 \equiv (\varphi_u^*\omega')^a{}_{b\mu} = \boldsymbol{\Lambda}'_{u\,c}{}^a\boldsymbol{\Lambda}'_u{}^d{}_b\omega'^c{}_{d\mu} + \boldsymbol{\Lambda}'_{u\,c}{}^a\partial_{\mu}\boldsymbol{\Lambda}'_u{}^c{}_b \equiv \boldsymbol{\Lambda}'_{u\,c}{}^a\partial_{\mu}\boldsymbol{\Lambda}'_u{}^c{}_b\,.
\end{equation}
with their infinitesimal versions~\eqref{eq:infisymcondgen} given by
\begin{equation}\label{eq:infisymcondwb}
(\mathcal{L}_{X_{\xi}}\theta')^a{}_{\mu} = -{\boldsymbol{\lambda}'}_{\xi}^a{}_b\theta'^b{}_{\mu}\,, \quad
0 \equiv (\mathcal{L}_{X_{\xi}}\omega')^a{}_{b\mu} = \partial_{\mu}{\boldsymbol{\lambda}'}_{\xi}^a{}_b\,.
\end{equation}
The second condition now simply states that the local Lie group homomorphism \(\boldsymbol{\Lambda}'_u\), and hence also the corresponding local Lie algebra homomorphism \({\boldsymbol{\lambda}'}_{\xi}\), must not depend on the spacetime point, and so instead of a local Lie group homomorphism one has a global one. Since for a given symmetry group there is only a limited number of such Lie group homomorphisms, this greatly simplifies the task of finding the symmetric tetrads. One simply has to choose a homomorphism \(\boldsymbol{\Lambda}': G \to \mathrm{SO}(1,3)\), and then solve the condition on \(\theta'^a{}_{\mu}\).

The aforementioned considerations also allow for a classification of symmetric teleparallel geometries. If \(\theta'^a{}_{\mu}\) satisfies the symmetry condition~\eqref{eq:infisymcondwb} for \(\boldsymbol{\Lambda}'\) in the Weitzenböck gauge, then \(\bar{\theta}'^a{}_{\mu} = \Lambda^a{}_b\theta'^b{}_{\mu}\) with a global Lorentz transformation \(\Lambda \in \mathrm{SO}(1,3)\) satisfies the corresponding symmetry condition for
\begin{equation}
\boldsymbol{\bar{\Lambda}}'_{u\,b}{}^a = \Lambda_b{}^d\Lambda^a{}_c\boldsymbol{\Lambda}'_{u\,d}{}^c\,.
\end{equation}
The two global homomorphisms \(\boldsymbol{\Lambda}', \boldsymbol{\bar{\Lambda}}': G \to \mathrm{SO}(1,3)\) can be regarded as isomorphic representations of \(G\) in the sense that they are equal up to conjugation with \(\Lambda\). This leads us to the following statement.

\begin{stmt}
When working in the Weitzenböck gauge, symmetric tetrads which belong to conjugated homomorphisms \(\boldsymbol{\Lambda}', \boldsymbol{\bar{\Lambda}}': G \to \mathrm{SO}(1,3)\) are related by global Lorentz transformations.
\end{stmt}

In the remainder of this article we will drop the prime (\('\)) and work in the Weitzenböck gauge whenever we are determining symmetric tetrads. Instead we will use the prime when we transform the tetrads we have found into a (sometimes more convenient) non-Weitzenböck gauge.

\subsection{Field equations for symmetric spacetimes}\label{ssec:symfeq}

We finally address the question how the notion of symmetry developed in the preceding sections is of use for the problem of solving or simplifying the field equations of teleparallel gravity theories. Recall that for any teleparallel gravity theory with local Lorentz invariance the field equations can be written in the generic form \(E_{\mu\nu} = \theta^a{}_\mu \theta^b{}_\nu E_{ab} = \kappa\Theta_{\mu\nu}\), where the energy-momentum tensor \(\Theta_{\mu\nu}\) is symmetric as a consequence of local Lorentz invariance. In this form the Euler-Lagrange expression \(E_{\mu\nu}\) can be regarded as a tensor constructed from the metric, seen as function of the tetrads, the torsion of the teleparallel connection and the Levi-Civita covariant derivative defined by the metric, see section \ref{ssec:tpgrav}. If one imposes the symmetry conditions \(\varphi_u^*g = g\) and \(\varphi_u^*\Gamma = \Gamma\) of the metric and the connection, then one easily derives that also the Levi-Civita connection and the torsion satisfy the corresponding symmetry conditions, i.e., they are invariant under the action of the symmetry group. The same holds for their covariant derivatives, products and contractions. It thus follows that any tensor constructed from these quantities has the same property. This leads us to the following statement:

\begin{stmt}\label{stmt:symfeq}
	Let \(E_{\mu\nu}\) be an arbitrary tensor field constructed from the metric~\eqref{eq:metric}, the torsion~\eqref{eq:torsion} and their covariant derivatives with respect to the Levi-Civita connection or the teleparallel connection~\eqref{eq:affconn}. If the tetrad and the spin connection satisfy the symmetry conditions~\eqref{eq:finisymcondgen}, then also \(E_{\mu\nu}\) satisfies the symmetry condition, \(\varphi_u^*E = E\) for all \(u \in G\).
\end{stmt}

This statement in particular applies to the Euler-Lagrange equations of any teleparallel gravity theory mentioned in section~\ref{ssec:tpgrav}. Since imposing a symmetry of the form \(\varphi_u^*E = E\) in general reduces the number of independent components of \(E_{\mu\nu}\) or restricts their coordinate dependence, it usually leads to a simplification of the corresponding field equations, up to fully solving them. However, note that the converse statement is not true: depending on the choice of the theory, also less symmetric teleparallel geometries may lead to Euler-Lagrange expressions which are trivially invariant under the action of the symmetry group. The most simple example is TEGR, where \(E_{\mu\nu}\) is the Einstein tensor, and thus depends only on the metric, so that one may relax the symmetry condition on the teleparallel connection~\eqref{eq:affconn}. However, we will not discuss such specific models in this work, and keep the theory and its field equations fully generic.

\section{Tetrads and spin connections obeying particular symmetries}\label{sec:solutions}
We now apply the formalism developed in the previous section to particular symmetry groups and their actions on a four-dimensional spacetime manifold \(M\). We do so in the order of increasing symmetry. As an illustrative example, we start with axial symmetry under the group \(\mathrm{SO}(2)\) in section~\ref{ssec:axial}. We then continue with spherical symmetry under the group \(\mathrm{SO}(3)\) in section~\ref{ssec:spherical}. In section~\ref{ssec:cosmo}, we discuss different types of cosmological symmetry, which are characterized by spatially homogeneous and isotropic spacetimes. Finally, we come to maximally symmetric spacetimes in section~\ref{ssec:maximal}.

\subsection{$\mathrm{SO}(2)$: axially symmetric spacetime}\label{ssec:axial}
We start our discussion of particular symmetries with an illustrative example. For this purpose we discuss the symmetry group \(\mathrm{SO}(2)\) of an axially symmetric spacetime. We use spherical coordinates \((t, r, \vartheta, \varphi)\), as it will turn out to be convenient later, when we discuss larger symmetry groups. In these coordinates the single generator of rotations around the polar axis is given by
\begin{equation}\label{eq:genvecz}
X_z = -\partial_{\varphi}\,.
\end{equation}
In order to construct a symmetric tetrad, we now need to choose a local homomorphism \(\boldsymbol{\Lambda}: \mathrm{SO}(2) \times M \to \mathrm{SO}(1,3)\). However, recall that we can simplify this choice by working in the Weitzenböck gauge \(\omega^a{}_{b\mu} = 0\), in which one of the symmetry conditions becomes equal to the condition that \(\boldsymbol{\Lambda}\) is constant over spacetime \(M\), and hence simply constitutes a (global) homomorphism \(\boldsymbol{\Lambda}: \mathrm{SO}(2) \to \mathrm{SO}(1,3)\). Of course, there are numerous possibilities to make this choice. Here we restrict ourselves to a simple example, and choose
\begin{equation}\label{eq:axgrouphomo}
\boldsymbol{\Lambda}(R(\phi)) = \begin{pmatrix}
1 & 0 & 0 & 0\\
0 & \cos\phi & -\sin\phi & 0\\
0 & \sin\phi & \cos\phi & 0\\
0 & 0 & 0 & 1
\end{pmatrix}\,,
\end{equation}
where \(R(\phi) \in \mathrm{SO}(2)\) denotes a rotation which is parametrized by an arbitrary angle \(\phi \in [0, 2\pi)\). The differential \(\boldsymbol{\lambda}\)~\eqref{eq:lochomo} of \(\boldsymbol{\Lambda}\) acts on the generator of rotation as
\begin{equation}\label{eq:axalghomo}
X_z \mapsto \begin{pmatrix}
0 & 0 & 0 & 0\\
0 & 0 & -1 & 0\\
0 & 1 & 0 & 0\\
0 & 0 & 0 & 0
\end{pmatrix}\,.
\end{equation}
With this assignment, we can now explicitly write out the first equation of the symmetry condition~\eqref{eq:infisymcondwb}, which reads
\begin{equation}
0 = (\mathcal{L}_{X_z}\theta)^a{}_{\mu} + \boldsymbol{\lambda}_z^a{}_b\theta^b{}_{\mu} = \begin{pmatrix}
-\partial_{\varphi}\theta^0{}_{\mu}\\
-\partial_{\varphi}\theta^1{}_{\mu}\\
-\partial_{\varphi}\theta^2{}_{\mu}\\
-\partial_{\varphi}\theta^3{}_{\mu}
\end{pmatrix} + \begin{pmatrix}
0 & 0 & 0 & 0\\
0 & 0 & -1 & 0\\
0 & 1 & 0 & 0\\
0 & 0 & 0 & 0
\end{pmatrix} \cdot \begin{pmatrix}
\theta^0{}_{\mu}\\
\theta^1{}_{\mu}\\
\theta^2{}_{\mu}\\
\theta^3{}_{\mu}
\end{pmatrix} = \begin{pmatrix}
-\partial_{\varphi}\theta^0{}_{\mu}\\
-\partial_{\varphi}\theta^1{}_{\mu} - \theta^2{}_{\mu}\\
-\partial_{\varphi}\theta^2{}_{\mu} + \theta^1{}_{\mu}\\
-\partial_{\varphi}\theta^3{}_{\mu}
\end{pmatrix}\,.
\end{equation}
From these equations one now immediately finds that the tetrad must be of the form
\begin{equation}\label{eq:axtetradwb}
\theta^0{}_{\mu} = C^0{}_{\mu}\,, \quad
\theta^1{}_{\mu} = C^1{}_{\mu}\cos\varphi - C^2{}_{\mu}\sin\varphi\,, \quad
\theta^2{}_{\mu} = C^1{}_{\mu}\sin\varphi + C^2{}_{\mu}\cos\varphi\,, \quad
\theta^3{}_{\mu} = C^3{}_{\mu}\,,
\end{equation}
where the 16 functions \(C^a{}_{\mu}\) depend only on the remaining coordinates \(t, r, \vartheta\). Alternatively, one may perform a local Lorentz transformation to obtain the tetrad
\begin{equation}\label{eq:axtetradconst}
\theta'^a{}_{\mu} = \Lambda^a{}_b\theta^b{}_{\mu} = C^a{}_{\mu}\,,
\end{equation}
whose components do not depend on the spacetime coordinate \(\varphi\) at all, where the Lorentz transformation is given by
\begin{equation}
\Lambda^a{}_b = \begin{pmatrix}
1 & 0 & 0 & 0\\
0 & \cos\varphi & \sin\varphi & 0\\
0 & -\sin\varphi & \cos\varphi & 0\\
0 & 0 & 0 & 1
\end{pmatrix}\,,
\end{equation}
and is hence given by a rotation whose angle agrees with the azimuth angle \(\varphi\) in the spherical coordinate system we use. In this case, however, one must include a non-trivial spin connection, since the local Lorentz transformation leads to a non-Weitzenböck gauge. The non-vanishing spin connection components read
\begin{equation}\label{eq:scaxsymm}
\omega^1{}_{2\varphi} = -\omega^2{}_{1\varphi} = -1\,.
\end{equation}
 The tetrad \eqref{eq:axtetradwb} or equivalently the tetrad \eqref{eq:axtetradconst} with spin connection \eqref{eq:scaxsymm} represent the most general teleparallel geometry which has axial symmetry implemented by the group homomorphism~\eqref{eq:axgrouphomo}. Observe that by statement~\ref{stmt:symfeq} of section \ref{ssec:symfeq} this implies that the field equations for any teleparallel theory of gravity also obey the symmetry conditions. Hence, in particular that the anti-symmetric part of the field equations can not depend on $\phi$ as we show in \eqref{eq:Eaxial}. This can also be seen by explicitly calculating the metric~\eqref{eq:metric} and affine connection~\eqref{eq:affconn}, which read
\begin{equation}
g_{\mu\nu} = \eta_{ab}C^a{}_{\mu}C^b{}_{\nu}
\end{equation}
and
\begin{equation}
\Gamma^{\mu}{}_{\nu\rho} = \begin{cases}
C_2{}^{\mu}C^1{}_{\nu} - C_1{}^{\mu}C^2{}_{\nu} & \text{for $\rho = \varphi$,}\\
C_a{}^{\mu}\partial_{\rho}C^a{}_{\nu} & \text{otherwise,}
\end{cases}
\end{equation}
where \(C_a{}^{\mu}\) denotes the inverse of \(C^a{}_{\mu}\). Note that these do not depend on the gauge choice, and are thus equivalently obtained from \(\theta^a{}_{\mu}\) given by~\eqref{eq:axtetradwb} in the Weitzenböck gauge or by the corresponding non-Weitzenböck expressions \(\theta'^a{}_{\mu}\) and \(\omega'^a{}_{b\mu}\). It also immediately follows that their Lie derivatives~\eqref{eq:metsymcond} and~\eqref{eq:affsymcond} vanish,
\begin{equation}
(\mathcal{L}_{X_z}g)_{\mu\nu} = -\partial_{\varphi}g_{\mu\nu} = 0\,, \quad
(\mathcal{L}_{X_z}\Gamma)^{\mu}{}_{\nu\rho} = -\partial_{\varphi}\Gamma^{\mu}{}_{\nu\rho} = 0\,,
\end{equation}
since their components do not depend on the coordinate \(\varphi\). Hence, also any tensorial expression \(E_{\mu\nu}\) constructed from these two quantities only is independent of \(\varphi\), and therefore possesses axial symmetry. A special case of axial symmetric teleparallel geometry has been applied to $f(T,\phi)$ gravity in \cite{Jarv:2019ctf}.

Of course the teleparallel geometry we obtained depends on the choice of homomorphism \(\boldsymbol{\Lambda}\). This can be seen explicitly by choosing it differently, such as the trivial homomorphism
\begin{equation}
\boldsymbol{\bar{\Lambda}}: \mathrm{SO}(2) \to \mathrm{SO}(1,3), U \mapsto \mathds{1}\,.
\end{equation}
In this case the condition on the corresponding tetrad \(\bar{\theta}^a{}_{\mu}\) in the Weitzenböck gauge simply reads
\begin{equation}
0 = (\mathcal{L}_{X_z}\bar{\theta})^a{}_{\mu} = -\partial_{\varphi}\bar{\theta}^a{}_{\mu}\,,
\end{equation}
which is solved by \(\bar{\theta}^a{}_{\mu} = C^a{}_{\mu}\) with the same free functions \(C^a{}_{\mu}\) of \(t, r, \vartheta\) as above. Note that this solution differs from the previous solution, \(\bar{\theta}^a{}_{\mu} \neq \theta^a{}_{\mu}\), as its components are independent of \(\varphi\) already in the Weitzenböck gauge. This can also be seen by deriving the metric and affine connection. While the metric remains the same, \(\bar{g}_{\mu\nu} = g_{\mu\nu}\), the connection now becomes \(\bar{\Gamma}^{\mu}{}_{\nu\rho} = C_a{}^{\mu}\partial_{\rho}C^a{}_{\nu} \neq \Gamma^{\mu}{}_{\nu\rho}\). The latter implies that also tensorial field equations derived from these quantities differ, and so it is indeed an inequivalent solution. Nevertheless, also \(\bar{\Gamma}^{\mu}{}_{\nu\rho}\) satisfies the symmetry condition~\eqref{eq:affsymcond}.

Note that by imposing axial symmetry we have completely fixed the dependence on the coordinate \(\varphi\). Nevertheless, the obtained tetrads are still very generic. They serve as an illustrative example and as ansatz to solve field equations of teleparallel theories of gravity. Moreover, they are the starting point for constructing tetrads with higher symmetries. This will be done in the following sections.

\subsection{$\mathrm{SO}(3)$: spherically symmetric spacetime}\label{ssec:spherical}
In the next step we extend the axial symmetry discussed in the previous section and consider spherical symmetry. Due to the length of the results we split the presentation in two parts. First we present the most general spherically symmetric tetrad in Weitzenböck gauge, as well as the corresponding metric and affine connection. Afterwards we apply Lorentz transformations to simplify the tetrad on the cost of a non-vanishing spin connection. By deriving again the metric and the affine connection from the transformed tetrad and spin connection we demonstrate explicitly the local Lorentz invariance of these spacetime structures.

\subsubsection{The $\mathrm{SO}(3)$ tetrad in Weitzenböck gauge}
In the next step we extend the axial symmetry discussed in the previous section and consider spherical symmetry. This can be achieved by introducing two additional symmetry generating vector fields, which take the form
\begin{equation}\label{eq:genvecxy}
X_x = \sin\varphi\partial_{\vartheta} + \frac{\cos\varphi}{\tan\vartheta}\partial_{\varphi}\,, \quad
X_y = -\cos\varphi\partial_{\vartheta} + \frac{\sin\varphi}{\tan\vartheta}\partial_{\varphi}\,.
\end{equation}
in the spherical coordinates we have chosen. In order to construct symmetric tetrads, we again need to choose a homomorphism \(\boldsymbol{\Lambda}\), but now with the domain given by the rotation group \(\mathrm{SO}(3)\). A simple and canonical choice is given by realizing that \(\mathrm{SO}(3)\) is embedded in \(\mathrm{SO}(1,3)\) via the homomorphism
\begin{equation}\label{eq:rotgrouphomo}
\boldsymbol{\Lambda}: \mathrm{SO}(3) \to \mathrm{SO}(1,3), U \mapsto \begin{pmatrix}
1 & 0\\
0 & U\\
\end{pmatrix}\,.
\end{equation}
Also note that by restriction to rotations around the polar axis, one obtains the homomorphism~\eqref{eq:axgrouphomo} discussed before. Hence, the corresponding Lie algebra homomorphism~\eqref{eq:axalghomo} is simply extended by the two additional assignments
\begin{equation}\label{eq:rotalghomo}
X_x \mapsto \begin{pmatrix}
0 & 0 & 0 & 0\\
0 & 0 & 0 & 0\\
0 & 0 & 0 & -1\\
0 & 0 & 1 & 0
\end{pmatrix}\,, \quad
X_y \mapsto \begin{pmatrix}
0 & 0 & 0 & 0\\
0 & 0 & 0 & 1\\
0 & 0 & 0 & 0\\
0 & -1 & 0 & 0
\end{pmatrix}\,.
\end{equation}
In order to solve the symmetry conditions, it is helpful to proceed in two steps. In the first step one makes use of the already determined axially symmetric tetrad~\eqref{eq:axtetradwb} and imposes the condition
\begin{equation}
\sin\varphi\left[(\mathcal{L}_{X_y}\theta)^a{}_{\mu} + \boldsymbol{\lambda}_y^a{}_b\theta^b{}_{\mu}\right] + \cos\varphi\left[(\mathcal{L}_{X_x}\theta)^a{}_{\mu} + \boldsymbol{\lambda}_x^a{}_b\theta^b{}_{\mu}\right] = 0\,,
\end{equation}
which is a linear combination of the symmetry conditions to be imposed, and has the advantage that the resulting equations are essentially decoupled. Further, one finds that some of the resulting equations are satisfied identically, while others do not contain derivatives, and so yield algebraic equations for the components \(C^a{}_{\mu}\) in the axially symmetric tetrad~\eqref{eq:axtetradwb}. First, one obtains the equations
\begin{equation}
C^0{}_{\vartheta} = C^0{}_{\varphi} = C^2{}_t = C^2{}_r = 0\,,
\end{equation}
so that these components must vanish. After imposing these conditions one further finds the equations
\begin{equation}
C^1{}_t\cos\vartheta - C^3{}_t\sin\vartheta = C^1{}_r\cos\vartheta - C^3{}_r\sin\vartheta = C^2{}_{\vartheta}\sin^2\vartheta - C^3{}_{\varphi} = C^2{}_{\varphi} + C^3{}_{\vartheta} = 0\,,
\end{equation}
which allow eliminating four further components. Finally, one finds the conditions
\begin{equation}
C^1{}_{\varphi} + C^2{}_{\vartheta}\cos\vartheta\sin\vartheta = C^1{}_{\vartheta}\sin\vartheta + C^3{}_{\vartheta}\cos\vartheta = 0\,.
\end{equation}
In total one thus finds 10 algebraic equations of the 16 components \(C^a{}_{\mu}\). The most general solution of these equations can be expressed in terms of 6 free functions \(C_1, \ldots, C_6\) of the coordinates \(t, r, \vartheta\) in the form
\begin{subequations}\label{eq:sphertetradwb}
\begin{align}
\theta^0 &= C_1\dd t + C_2\dd r\,,\\
\theta^1 &= \sin\vartheta\cos\varphi(C_3\dd t + C_4\dd r) + (C_5\cos\vartheta\cos\varphi - C_6\sin\varphi)\dd\vartheta - \sin\vartheta(C_5\sin\varphi + C_6\cos\vartheta\cos\varphi)\dd\varphi\,,\\
\theta^2 &= \sin\vartheta\sin\varphi(C_3\dd t + C_4\dd r) + (C_5\cos\vartheta\sin\varphi + C_6\cos\varphi)\dd\vartheta + \sin\vartheta(C_5\cos\varphi - C_6\cos\vartheta\sin\varphi)\dd\varphi\,,\\
\theta^3 &= \cos\vartheta(C_3\dd t + C_4\dd r) - C_5\sin\vartheta\dd\vartheta + C_6\sin^2\vartheta\dd\varphi\,.
\end{align}
\end{subequations}
We then come to the second step, and impose the linear combination
\begin{equation}
\cos\varphi\left[(\mathcal{L}_{X_y}\theta)^a{}_{\mu} + \boldsymbol{\lambda}_y^a{}_b\theta^b{}_{\mu}\right] - \sin\varphi\left[(\mathcal{L}_{X_x}\theta)^a{}_{\mu} + \boldsymbol{\lambda}_x^a{}_b\theta^b{}_{\mu}\right] = 0
\end{equation}
on the tetrad~\eqref{eq:sphertetradwb}. The resulting equations now simply read
\begin{equation}
\partial_{\vartheta}C_1 = \ldots = \partial_{\vartheta}C_6 = 0\,,
\end{equation}
which means that the functions \(C_1, \ldots, C_6\) may depend only on \(t\) and \(r\), so that the dependence on \(\vartheta\) is fully determined. The tetrad~\eqref{eq:sphertetradwb} is then the most general spherically symmetric tetrad in the Weitzenböck gauge. Its corresponding metric \eqref{eq:metric} has non-vanishing components
\begin{align}\label{eq:metspher1}
	g_{tt} = C_3^2 - C_1^2\,, \quad
	g_{rr} = C_4^2 - C_2^2\,, \quad
	g_{rt} = g_{tr} = C_3C_4 - C_1C_2\,, \quad
	g_{\vartheta\vartheta} = C_5^2 + C_6^2\,, \quad
	g_{\varphi\varphi} = g_{\vartheta\vartheta}\sin^2\vartheta\,,
\end{align}
while the corresponding affine teleparallel connection \eqref{eq:affconn} is given by the non-vanishing components
\begin{align}\label{eq:affspher}
\Gamma^{t}{}_{tt} 					&= \frac{C_4 \partial_t C_1 - C_2 \partial_t C_3}{C_1 C_4 - C_2 C_3} &
\Gamma^{r}{}_{tt} 					&= \frac{C_1 \partial_t C_3 - C_3\partial_t C_1}{C_1 C_4 - C_2 C_3} &
\Gamma^{\vartheta}{}_{t\vartheta} 	&= \frac{C_3 C_5}{C_5^2 + C_6^2} &
\Gamma^{\varphi}{}_{t\vartheta} 	&= \frac{-1}{\sin^2\vartheta}\Gamma^{\vartheta}{}_{t\varphi} \nonumber\\
\Gamma^{t}{}_{tr} 					&= \frac{C_4 \partial_r C_1 - C_2 \partial_r C_3}{C_1 C_4 - C_2 C_3} &
\Gamma^{r}{}_{tr} 					&= \frac{C_1 \partial_r C_3 - C_3\partial_r C_1}{C_1 C_4 - C_2 C_3} &
\Gamma^{\vartheta}{}_{t\varphi} 	&= \frac{C_3 C_6 \sin\vartheta}{C_5^2 + C_6^2} &
\Gamma^{\varphi}{}_{t\varphi} 		&= \Gamma^{\vartheta}{}_{t\vartheta}\nonumber\\
\Gamma^{t}{}_{rt} 					&= \frac{C_4 \partial_t C_2 - C_2 \partial_t C_4}{C_1 C_4 - C_2 C_3} &
\Gamma^{r}{}_{rt} 					&= \frac{C_1 \partial_t C_4 - C_3\partial_t C_2}{C_1 C_4 - C_2 C_3} &
\Gamma^{\vartheta}{}_{r\vartheta} 	&= \frac{C_4 C_5}{C_5^2 + C_6^2} &
\Gamma^{\varphi}{}_{r\vartheta} 	&= \frac{-1}{\sin^2\vartheta}\Gamma^{\vartheta}{}_{r\varphi} \nonumber\\
\Gamma^{t}{}_{rr} 					&= \frac{C_4 \partial_r C_2 - C_2 \partial_r C_4}{C_1 C_4 - C_2 C_3} &
\Gamma^{r}{}_{rr} 					&= \frac{C_1 \partial_r C_4 - C_3\partial_r C_2}{C_1 C_4 - C_2 C_3} &
\Gamma^{\vartheta}{}_{r\varphi} 	&= \frac{C_4 C_6 \sin\vartheta}{C_5^2 + C_6^2} &
\Gamma^{\varphi}{}_{r\varphi} 		&= \Gamma^{\vartheta}{}_{r\vartheta} \nonumber\\
\Gamma^{t}{}_{\vartheta\vartheta} 	&= \frac{C_2 C_5}{C_1 C_4 - C_2 C_3} &
\Gamma^{r}{}_{\vartheta\vartheta} 	&= \frac{- C_1 C_5}{C_1 C_4 - C_2 C_3} &
\Gamma^{\vartheta}{}_{\vartheta t} 	&= \frac{C_5 \partial_t C_5 + C_6 \partial_t C_6}{C_5^2 + C_6^2} &
\Gamma^{\varphi}{}_{\vartheta t} 	&= - \frac{1}{\sin\vartheta^2}\Gamma^{\vartheta}{}_{\varphi t} \nonumber\\
\Gamma^{t}{}_{\vartheta\varphi} 	&= \frac{C_2 C_6 \sin\vartheta}{C_1 C_4 - C_2 C_3} &
\Gamma^{r}{}_{\vartheta\varphi} 	&= \frac{-C_1 C_6 \sin\vartheta}{C_1 C_4 - C_2 C_3} &
\Gamma^{\vartheta}{}_{\vartheta r} 	&= \frac{C_5 \partial_r C_5 + C_6 \partial_r C_6}{C_5^2 + C_6^2} &
\Gamma^{\varphi}{}_{\vartheta r} 	&= - \frac{1}{\sin\vartheta^2}\Gamma^{\vartheta}{}_{\varphi r} \nonumber\\
\Gamma^{t}{}_{\varphi\vartheta} 	&= -\Gamma^{t}{}_{\vartheta\varphi} &
\Gamma^{r}{}_{\varphi\vartheta} 	&= -\Gamma^{r}{}_{\vartheta\varphi} &
\Gamma^{\vartheta}{}_{\varphi t} 	&= \frac{\sin\vartheta (C_6 \partial_t C_5 - C_5 \partial_t C_6)}{C_5^2 + C_6^2} &
\Gamma^{\varphi}{}_{\varphi t} 		&= \Gamma^{\vartheta}{}_{\vartheta t} &\nonumber\\
\Gamma^{t}{}_{\varphi\varphi} 		&= \sin^2\vartheta \Gamma^{t}{}_{\vartheta\vartheta} &
\Gamma^{r}{}_{\varphi\varphi} 		&= \sin^2\vartheta \Gamma^{r}{}_{\vartheta\vartheta} &
\Gamma^{\vartheta}{}_{\varphi r} 	&= \frac{\sin\vartheta(C_6\partial_r C_5 - C_5 \partial_r C_6)}{C_5^2 + C_6^2} &
\Gamma^{\varphi}{}_{\varphi r} 		&= \Gamma^{\vartheta}{}_{\vartheta r} \nonumber\\
& &
& &
\Gamma^{\vartheta}{}_{\varphi\varphi} &= - \cos\vartheta \sin\vartheta &
\Gamma^{\varphi}{}_{\vartheta\varphi} &= \Gamma^{\varphi}{}_{\varphi\vartheta} = \cot\vartheta\,.
\end{align}
It is now easy to check that the metric and connection satisfy the spherical symmetry conditions, i.e., their Lie derivatives~\eqref{eq:metsymcond} and~\eqref{eq:affsymcond} vanish for the spherical symmetry generators.

Observe again that by statement~\ref{stmt:symfeq} of section \ref{ssec:symfeq} the symmetry of the tetrad implies that the field equations for any teleparallel theory of gravity also obey the symmetry conditions. Hence, in particular the only non-vanishing components of the anti-symmetric part of the field equations are $E_{[tr]}$ and $E_{[\vartheta\phi]}$, as we show in \eqref{eq:Espherical}.

\subsubsection{The block diagonal $\mathrm{SO}(3)$ tetrad and its spin connection}

Simpler forms of the spherically symmetric tetrad can be obtained by applying Lorentz transformations to \eqref{eq:sphertetradwb}. In consequence, the spin connection corresponding to the new tetrads will be non-vanishing, which ensures local Lorentz invariance on spacetime. We demonstrate the latter feature explicitly by calculating the metric and the affine connection again after we applied several local Lorentz transformations. In the following three non-Weitzenböck tetrads will be constructed.

We begin with the local Lorentz transformation given by
\begin{equation}\label{eq:spherlt1}
\Lambda^a{}_b = \begin{pmatrix}
1 & 0 & 0 & 0\\
0 & \sin\vartheta\cos\varphi & \sin\vartheta\sin\varphi & \cos\vartheta\\
0 & \cos\vartheta\cos\varphi & \cos\vartheta\sin\varphi & -\sin\vartheta\\
0 & -\sin\varphi & \cos\varphi & 0
\end{pmatrix}\,.
\end{equation}
It yields the new tetrad
\begin{equation}\label{eq:tetspher1}
\theta'^0 = C_1\dd t + C_2\dd r\,, \quad
\theta'^1 = C_3\dd t + C_4\dd r\,, \quad
\theta'^2 = C_5\dd\vartheta - C_6\sin\vartheta\dd\varphi\,, \quad
\theta'^3 = C_6\dd\vartheta + C_5\sin\vartheta\dd\varphi\,,
\end{equation}
where we now have a non-trivial spin connection, whose non-vanishing components are given by
\begin{equation}\label{eq:spiconspher1}
\omega'^1{}_{2\vartheta} = -\omega'^2{}_{1\vartheta} = -1\,, \quad
\omega'^1{}_{3\varphi} = -\omega'^3{}_{1\varphi} = -\sin\vartheta\,, \quad
\omega'^2{}_{3\varphi} = -\omega'^3{}_{2\varphi} = -\cos\vartheta\,.
\end{equation}

In order to further simplify the tetrad, we replace \(C_5\) and \(C_6\) by the equivalent parametrization
\begin{equation}
C_5 = C_s\cos C_{\alpha}\,, \quad
C_6 = C_s\sin C_{\alpha}\,,
\end{equation}
where now \(C_s\) and \(C_{\alpha}\) are free functions of the coordinates \(t\) and \(r\). Performing the additional local Lorentz transformation
\begin{equation}\label{eq:spherlt2}
\Lambda'^a{}_b = \begin{pmatrix}
1 & 0 & 0 & 0\\
0 & 1 & 0 & 0\\
0 & 0 & \cos C_{\alpha} & \sin C_{\alpha}\\
0 & 0 & -\sin C_{\alpha} & \cos C_{\alpha}
\end{pmatrix}
\end{equation}
then yields the tetrad
\begin{equation}\label{eq:tetspher2}
\theta''^0 = C_1\dd t + C_2\dd r\,, \quad
\theta''^1 = C_3\dd t + C_4\dd r\,, \quad
\theta''^2 = C_s\dd\vartheta\,, \quad
\theta''^3 = C_s\sin\vartheta\dd\varphi\,,
\end{equation}
while the spin connection now has components
\begin{align}
\omega''^1{}_{2\vartheta} &= -\omega''^2{}_{1\vartheta} = -\cos C_{\alpha}\,,  &
\omega''^1{}_{2\varphi} &= -\omega''^2{}_{1\varphi} = -\sin\vartheta\sin C_{\alpha}\,, &
\omega''^1{}_{3\varphi} &= -\omega''^3{}_{1\varphi} = -\sin\vartheta\cos C_{\alpha}\,, \nonumber\\
\omega''^1{}_{3\vartheta} &= -\omega''^3{}_{1\vartheta} = \sin C_{\alpha}\,, &
\omega''^2{}_{3t} &= -\omega''^3{}_{2t} = -\partial_t C_{\alpha}\,, &
\omega''^2{}_{3r} &= -\omega''^3{}_{2r} = -\partial_r C_{\alpha}\,, \nonumber\\
\omega''^2{}_{3\varphi} &= -\omega''^3{}_{2\varphi} = -\cos\vartheta\,.\label{eq:spiconspher2}
\end{align}

Finally, one may use the equivalent parametrization defined through
\begin{equation}
C_1 = C_t\cosh C_{\upsilon}\,, \quad
C_2 = C_r\sinh C_{\psi}\,, \quad
C_3 = C_t\sinh C_{\upsilon}\,, \quad
C_4 = C_r\cosh C_{\psi}\,,
\end{equation}
where now \(C_t, C_r, C_{\upsilon}, C_{\psi}\) are free functions of \(t\) and \(r\). One may then apply one of the following two local Lorentz transformations. The first one is given by
\begin{equation}\label{eq:spherlt3}
\Lambda''^a{}_b = \begin{pmatrix}
\cosh C_{\upsilon} & -\sinh C_{\upsilon} & 0 & 0\\
-\sinh C_{\upsilon} & \cosh C_{\upsilon} & 0 & 0\\
0 & 0 & 1 & 0\\
0 & 0 & 0 & 1
\end{pmatrix}
\end{equation}
and yields the almost diagonal tetrad
\begin{equation}\label{eq:tetspher3a}
\theta'''^0 = C_t\dd t - C_r\sinh(C_{\upsilon} - C_{\psi})\dd r\,, \quad
\theta'''^1 = C_r\cosh(C_{\upsilon} - C_{\psi})\dd r\,, \quad
\theta'''^2 = C_s\dd\vartheta\,, \quad
\theta'''^3 = C_s\sin\vartheta\dd\varphi\,,
\end{equation}
together with the now more involved non-vanishing spin connection
\begin{align}
\omega'''^0{}_{2\varphi}    = \omega'''^2{}_{0\varphi} &= \sinh C_{\upsilon}\sin\vartheta\sin C_{\alpha}\,, &
\omega'''^0{}_{2\vartheta}  = \omega'''^2{}_{0\vartheta} &= \sinh C_{\upsilon}\cos C_{\alpha}\,,  &
\omega'''^0{}_{1t}          = \omega'''^1{}_{0t} &= \partial_t C_{\upsilon}\,, \nonumber\\
-\omega'''^1{}_{2\varphi}   = \omega'''^2{}_{1\varphi} &= \cosh C_{\upsilon}\sin\vartheta\sin C_{\alpha}\,,&
-\omega'''^1{}_{2\vartheta} = \omega'''^2{}_{1\vartheta} &= \cosh C_{\upsilon}\cos C_{\alpha}\,, &
-\omega'''^2{}_{3t}         = \omega'''^3{}_{2t} &= \partial_t C_{\alpha}\,, \nonumber\\
\omega'''^0{}_{3\varphi}    = \omega'''^3{}_{0\varphi} &= \sinh C_{\upsilon}\sin\vartheta\cos C_{\alpha} &
\omega'''^0{}_{3\vartheta}  = \omega'''^3{}_{0\vartheta} &= -\sinh C_{\upsilon}\sin C_{\alpha}\,, &
\omega'''^0{}_{1r}          = \omega'''^1{}_{0r} &= \partial_r C_{\upsilon}\,, \nonumber\\
-\omega'''^1{}_{3\varphi}   = \omega'''^3{}_{1\varphi} &= \cosh C_{\upsilon}\sin\vartheta\cos C_{\alpha}\,, &
-\omega'''^1{}_{3\vartheta} = \omega'''^3{}_{1\vartheta} &= -\cosh C_{\upsilon}\sin C_{\alpha}\,, &
-\omega'''^2{}_{3r}         = \omega'''^3{}_{2r} &= \partial_r C_{\alpha}\,, \nonumber\\&& &&
-\omega'''^2{}_{3\varphi}   = \omega'''^3{}_{2\varphi} &= \cos\vartheta\,. \label{eq:spiconspher3}
\end{align}


The second alternative is obtained by applying a local Lorentz transformation \(\tilde{\Lambda}''^a{}_b\) instead of \(\Lambda''^a{}_b\), which takes the same form~\eqref{eq:spherlt3}, but with \(C_{\psi}\) instead of \(C_{\upsilon}\). The resulting spin connection \(\tilde{\omega}''^a{}_{b\mu}\) takes the same form~\eqref{eq:spiconspher3} as \(\omega''^a{}_{b\mu}\), but with \(C_{\psi}\) instead of \(C_{\upsilon}\), while the tetrad becomes
\begin{equation}\label{eq:tetspher3b}
\tilde{\theta}'''^0 = C_t\cosh(C_{\upsilon} - C_{\psi})\dd t\,, \quad
\tilde{\theta}'''^1 = C_t\sinh(C_{\upsilon} - C_{\psi})\dd t + C_r\dd r\,, \quad
\tilde{\theta}'''^2 = C_s\dd\vartheta\,, \quad
\tilde{\theta}'''^3 = C_s\sin\vartheta\dd\varphi\,.
\end{equation}
Comparing this result to the original tetrad~\eqref{eq:sphertetradwb} in the Weitzenböck gauge, we see instead of the six free functions \(C_1, \ldots, C_6\) in the tetrad one now only has four free functions \(C_t, C_r, C_s, C_{\upsilon} - C_{\psi}\) in the tetrad, while the remaining two free functions \(C_{\alpha}\) and either \(C_{\upsilon}\) or \(C_{\psi}\) determine the spin connection. Note further that if \(C_{\upsilon}(t, r) = C_{\psi}(t, r)\) for all \(t, r\) we find that \(\tilde{\Lambda}''^a{}_b = \Lambda''^a{}_b\) and \(\tilde{\omega}'''^a{}_{b\mu} = \omega'''^a{}_{b\mu}\), while the tetrad \(\tilde{\theta}'''^a{}_{\mu} = \theta'''^a{}_{\mu}\) becomes diagonal. An example for this tetrad given by the parameter functions
\begin{equation}
C_1 = 1\,, \quad
C_2 = -\sqrt{\frac{\zeta}{\xi} - \zeta}\,, \quad
C_3 = -\sqrt{1 - \xi}\,, \quad
C_4 = \sqrt{\frac{\zeta}{\xi}}\,, \quad
C_5 = r\sqrt{\frac{\xi}{\zeta}}\,, \quad
C_6 = \mp r\sqrt{1 - \frac{\xi}{\zeta}}\,,
\end{equation}
and hence
\begin{equation}
C_t = \sqrt{\xi}\,, \quad
C_r = \sqrt{\zeta}\,, \quad
C_s = r\,, \quad
C_{\upsilon} = C_{\psi} = -\arcsinh\sqrt{\frac{1 - \xi}{\xi}}\,, \quad
C_{\alpha} = \mp\arcsin\sqrt{1 - \frac{\xi}{\zeta}}\,,
\end{equation}
where \(\xi\) and \(\zeta\) are functions of the radial coordinate \(r\) only, was presented in~\cite{Lin:2019tos}.

We also display a few quantities derived from this tetrad. First note that in order to be non-degenerate, its determinant
\begin{equation}
\det\theta = (C_1C_4 - C_2C_3)(C_5^2 + C_6^2)\sin\vartheta = C_tC_rC_s^2\cosh(C_{\upsilon} - C_{\psi})\sin\vartheta\,,
\end{equation}
and hence also the parameter functions \(C_t, C_r, C_s\), must be non-vanishing for all \(t, r\).

In the alternative parametrization we introduced above, the metric \eqref{eq:metspher1} becomes
\begin{equation}\label{eq:metspher2}
g_{tt} = -C_t^2\,, \quad
g_{rr} = C_r^2\,, \quad
g_{rt} = g_{tr} = C_tC_r\sinh(C_{\upsilon} - C_{\psi})\,, \quad
g_{\vartheta\vartheta} = C_s^2\,, \quad
g_{\varphi\varphi} = g_{\vartheta\vartheta}\sin^2\vartheta\,.
\end{equation}
Note that the metric depends only on the four functions \(C_t, C_r, C_s, C_{\upsilon} - C_{\psi}\) appearing in the tetrads~\eqref{eq:tetspher3a} and~\eqref{eq:tetspher3b}. The remaining functions, however, appear in the teleparallel connection~\eqref{eq:affconn}, which is more conveniently expressed in the latter parametrization. Its non-vanishing components then read, with $X=C_{\upsilon} - C_{\psi}$,
\begin{align}
\Gamma^t{}_{tt}		 					&= \frac{\partial_tC_t}{C_t} + \tanh(X)\partial_tC_{\upsilon} &
\Gamma^r{}_{tt} 						&= \frac{C_t\partial_tC_\upsilon}{C_r\cosh(X)}&
\Gamma^{\vartheta}{}_{t\vartheta} 		&= \frac{C_t\sinh C_{\upsilon}\cos C_{\alpha}}{C_s} &
\Gamma^{\varphi}{}_{t\vartheta} 		&= \frac{-1}{\sin^2\vartheta} \Gamma^{\vartheta}_{t\phi} & \nonumber\\
\Gamma^t{}_{tr} 						&= \frac{\partial_rC_t}{C_t} + \tanh(X)\partial_rC_{\upsilon} &
\Gamma^r{}_{tr} 						&= \frac{C_t\partial_rC_\upsilon}{C_r\cosh(X)} &
\Gamma^{\vartheta}{}_{t\varphi} 		&= \frac{C_t\sinh C_{\upsilon}\sin C_{\alpha}\sin\vartheta}{C_s}&
\Gamma^{\varphi}{}_{t\varphi}			&= \Gamma^{\vartheta}{}_{t\vartheta}\nonumber\\
\Gamma^t{}_{rt} 						&= \frac{C_r\partial_tC_{\psi}}{C_t\cosh(X)} &
\Gamma^r{}_{rt} 						&= \frac{\partial_tC_r}{C_r} - \tanh(X)\partial_tC_{\psi} &
\Gamma^{\vartheta}{}_{r\vartheta} 		&= \frac{C_r\cosh C_{\psi}\cos C_{\alpha}}{C_s} &
\Gamma^{\varphi}{}_{r\vartheta} 		&= \frac{-1}{\sin^2\vartheta}\Gamma^{\vartheta}{}_{r\varphi} &\nonumber\\
\Gamma^t{}_{rr} 						&= \frac{C_r\partial_rC_{\psi}}{C_t\cosh(X)} &
\Gamma^r{}_{rr} 						&= \frac{\partial_rC_r}{C_r} - \tanh(X)\partial_rC_{\psi} &
\Gamma^{\vartheta}{}_{r\varphi} 		&= \frac{C_r\cosh C_{\psi}\sin C_{\alpha}\sin\vartheta}{C_s}&
\Gamma^{\varphi}{}_{r\varphi} 			&=\Gamma^{\vartheta}{}_{r\vartheta} \nonumber\\
\Gamma^t{}_{\vartheta\vartheta} 		&= \frac{C_s\sinh C_{\psi}\cos C_{\alpha}}{C_t\cosh(X)} &
\Gamma^r{}_{\vartheta\vartheta} 		&= -\frac{C_s\cosh C_{\upsilon}\cos C_{\alpha}}{C_r\cosh(X)} &
\Gamma^{\vartheta}{}_{\vartheta t} 		&= \frac{\partial_tC_s}{C_s} &
\Gamma^{\varphi}{}_{\vartheta t} 		&= \frac{-1}{\sin^2\vartheta}\Gamma^{\vartheta}{}_{\varphi t}\nonumber\\
\Gamma^t{}_{\vartheta\varphi} 			&=  \frac{C_s\sinh C_{\psi}\sin C_{\alpha}\sin\vartheta}{C_t\cosh(X)} &
\Gamma^r{}_{\vartheta\varphi} 			&= -\frac{C_s\cosh C_{\upsilon}\sin C_{\alpha}\sin\vartheta}{C_r\cosh(X)}&
\Gamma^{\vartheta}{}_{\vartheta r} 		&= \Gamma^{\varphi}{}_{\varphi r} = \frac{\partial_rC_s}{C_s} &
\Gamma^{\varphi}{}_{\vartheta r} 		&= \frac{-1}{\sin^2\vartheta}\Gamma^{\vartheta}{}_{\varphi r} \nonumber\\
\Gamma^t{}_{\varphi\vartheta}			&= -\Gamma^t{}_{\vartheta\varphi}&
\Gamma^r{}_{\varphi\vartheta}			&= -\Gamma^r{}_{\vartheta\varphi} &
\Gamma^{\vartheta}{}_{\varphi t} 		&= -\partial_t C_{\alpha}\sin\vartheta &
\Gamma^{\varphi}{}_{\varphi t} 			&= \Gamma^{\vartheta}{}_{\vartheta t}\nonumber\\
\Gamma^t{}_{\varphi\varphi} 			&= \sin^2\vartheta\Gamma^t{}_{\vartheta\vartheta} &
\Gamma^r{}_{\varphi\varphi} 			&= \sin^2\vartheta\Gamma^r{}_{\vartheta\vartheta}&
\Gamma^{\vartheta}{}_{\varphi r} 		&= -\partial_r C_{\alpha}\sin\vartheta&
\Gamma^{\varphi}{}_{\varphi r} 			&= \Gamma^{\vartheta}{}_{\vartheta r}\nonumber\\
& &
& &
\Gamma^{\vartheta}{}_{\varphi\varphi} &= - \cos\vartheta \sin\vartheta &
\Gamma^{\varphi}{}_{\vartheta\varphi} &= \Gamma^{\varphi}{}_{\varphi\vartheta} = \cot\vartheta\,.\label{eq:connspher}
\end{align}

Again the metric and connection satisfy the spherical symmetry conditions, i.e., their Lie derivatives~\eqref{eq:metsymcond} and~\eqref{eq:affsymcond} vanish for the spherical symmetry generators. Comparing equations \eqref{eq:affspher} and \eqref{eq:connspher} demonstrates explicitly that the covariant formulation of teleparallel theories of gravity, which takes the spin connection into account properly, ensures local Lorentz invariance. The expressions of the affine connection are identical up to the redefinition of the functions $C_I$.

We also remark that although the most general symmetric tetrad we obtained cannot be brought into diagonal form by applying local Lorentz transformation, but retains an off-diagonal component as seen in the expressions~\eqref{eq:tetspher3a} and~\eqref{eq:tetspher3b}, which is related to the non-vanishing off-diagonal metric component \(g_{tr}\) in the metric~\eqref{eq:metspher1} and~\eqref{eq:metspher2}. One may of course achieve this diagonal form by a coordinate transformation \((t, r) \mapsto (\tilde{t}, \tilde{r})\), which does not influence the spherical symmetry, since it commutes with rotations. Under this transformation the tetrad retains its general form with six free parameter functions, but these functions transform according to
\begin{equation}
\tilde{C}_{1,3}(\tilde{t}, \tilde{r}) = C_{1,3}(t, r)\frac{\partial t}{\partial\tilde{t}} + C_{2,4}(t, r)\frac{\partial r}{\partial\tilde{t}}\,, \quad
\tilde{C}_{2,4}(\tilde{t}, \tilde{r}) = C_{1,3}(t, r)\frac{\partial t}{\partial\tilde{r}} + C_{2,4}(t, r)\frac{\partial r}{\partial\tilde{r}}\,, \quad
\tilde{C}_{5,6}(\tilde{t}, \tilde{r}) = C_{5,6}(t, r)\,,
\end{equation}
where \(t\) and \(r\) are now seen as functions of \(\tilde{t}\) and \(\tilde{r}\). It is thus possible to choose a transformation such that \(\tilde{C}_3\tilde{C}_4 - \tilde{C}_1\tilde{C}_2 = 0\).

Before we continue to cosmological symmetry, we'd like to to point out that the whole procedure of finding the spherically symmetric teleparallel geometry is based on the choice of the group homomorphism~\eqref{eq:rotgrouphomo}. One may of course pose the question whether it is necessary to pick a non-trivial group homomorphism, or whether one could proceed as shown for axial symmetry at the end of section~\ref{ssec:axial} and simply use the trivial homomorphism
\begin{equation}
\boldsymbol{\bar{\Lambda}}: \mathrm{SO}(3) \to \mathrm{SO}(1,3), U \mapsto \mathds{1}\,.
\end{equation}
Assume that \(\bar{\theta}\) is a symmetric tetrad in the Weitzenböck gauge, so that it satisfies the condition~\eqref{eq:finisymcondwb}, which in this case reduces to \((\varphi_u^*\bar{\theta})^a{}_{\mu} = \bar{\theta}^a{}_{\mu}\). Now consider a point \(x \in M\) at coordinates \((t, r, \vartheta, \varphi)\) with \(r > 0\). The orbit \(O\) of the group action of \(\mathrm{SO}(3)\) through \(x\) is a sphere \(S^2\) with fixed coordinates \((t, r)\). At the point \(x\), choose a tangent vector \(v = v^{\vartheta}\partial_{\vartheta} + v^{\varphi}\partial_{\varphi} = v^a\bar{e}_a(x)\) to this orbit, which is not the null vector. This implies that its components \(v^a = \bar{\theta}^a(v) = \bar{\theta}^a{}_{\vartheta}(x)v^{\vartheta} + \bar{\theta}^a{}_{\varphi}(x)v^{\varphi}\) in the basis dual to the tetrad \(\bar{\theta}\) are not all zero. Now for each \(x' \in O\) we can find some \(u \in G\) for which \(x' = \varphi_u(x)\). Then \(v' = \varphi_{u*}(v) = v^a\varphi_u(\bar{e}_a(x)) = v^a\bar{e}_a(x')\) would again be tangent to \(O\) and non-vanishing. Doing this for all \(x' \in O\) would yield a nowhere vanishing tangent vector field of \(O \cong S^2\). However, this is not possible, since the sphere is not parallelizable. Hence, no such tetrad field \(\bar{\theta}\) can exist. With this we demonstrated that the homomorphism between the symmetry group and the Lorentz group must be a non-trivial one to obtain a symmetric tetrad. We also remark that similar topological obstructions exist for the existence of ``good'' or ``proper'' tetrad in certain generalized teleparallel theories~\cite{Bejarano:2019fii}.

\subsection{Cosmological symmetry}\label{ssec:cosmo}
We now come to the class of cosmological symmetries. This class has the peculiar property that using the method we present in this article, it is possible to solve the antisymmetric part of the field equations of any teleparallel gravity theory, independent of its Lagrangian. This will be explained in section~\ref{sssec:cosmogen}. We then present the solutions for the three specific symmetries, corresponding to the spatial curvature parameter \(k = -1, 0, 1\). The flat case \(k = 0\) is discussed in section~\ref{sssec:cosmoflat}, while we discuss positive spatial curvature \(k = 1\) in section~\ref{sssec:cosmopos} and negative spatial curvature \(k = -1\) in section~\ref{sssec:cosmoneg}.

\subsubsection{General considerations}\label{sssec:cosmogen}
We start our discussion of cosmological symmetry with some general results which hold for any value of the spatial curvature parameter \(k\). Note that in this case we have in addition to the generating vector fields~\eqref{eq:genvecz} and~\eqref{eq:genvecxy} of rotations also the translation generators
\begin{subequations}\label{eq:genvectrans}
\begin{align}
X_1 &= \chi\sin\vartheta\cos\varphi\partial_r + \frac{\chi}{r}\cos\vartheta\cos\varphi\partial_{\vartheta} - \frac{\chi\sin\varphi}{r\sin\vartheta}\partial_{\varphi}\,,\\
X_2 &= \chi\sin\vartheta\sin\varphi\partial_r + \frac{\chi}{r}\cos\vartheta\sin\varphi\partial_{\vartheta} + \frac{\chi\cos\varphi}{r\sin\vartheta}\partial_{\varphi}\,,\\
X_3 &= \chi\cos\vartheta\partial_r - \frac{\chi}{r}\sin\vartheta\partial_{\vartheta}\,,
\end{align}
\end{subequations}
where we introduced the abbreviation \(\chi = \sqrt{1 - kr^2}\), and where we used spherical coordinates \((t, r, \vartheta, \varphi)\). Due to its high amount of symmetry this class of symmetries is of particular interest, since it poses strong restrictions on the field equations. Recall from section~\ref{ssec:symfeq} statement~\ref{stmt:symfeq}, that if the metric \(g_{\mu\nu}\) and the affine connection \(\Gamma^{\rho}{}_{\mu\nu}\) are invariant under the flow of a vector field \(X\), then also any Euler-Lagrange tensor \(E_{\mu\nu}\) constructed from these quantities is invariant. This in particular then also implies that its antisymmetric part \(E_{[\mu\nu]}\) is invariant. In the case of cosmological symmetry this is a very strong condition, due to the following statement:

\begin{stmt}
Any antisymmetric two-tensor with cosmological symmetry generated by the vector fields~\eqref{eq:genvecz}, \eqref{eq:genvecxy} and~\eqref{eq:genvectrans} vanishes.
\end{stmt}
\begin{proof}
This can simply be proven as follows. We start with the vector field \(X_z\) according to its definition~\eqref{eq:genvecz}. Invariance implies
\begin{equation}\label{eq:Eaxial}
0 = (\mathcal{L}_{X_z}E)_{[\mu\nu]} = \partial_{\varphi}E_{[\mu\nu]}\,.
\end{equation}
Hence, the components \(E_{[\mu\nu]}\) do not depend on \(\varphi\). We then continue with the generators~\eqref{eq:genvecxy}. A direct calculation shows that
\begin{equation}
(\cos\varphi\mathcal{L}_{X_x}E + \sin\varphi\mathcal{L}_{X_y}E)_{[\mu\nu]} = \begin{pmatrix}
0 & 0 & -\frac{E_{[t\varphi]}}{\sin^2\vartheta} & E_{[t\vartheta]}\\
0 & 0 & -\frac{E_{[r\varphi]}}{\sin^2\vartheta} & E_{[r\vartheta]}\\
\frac{E_{[t\varphi]}}{\sin^2\vartheta} & \frac{E_{[r\varphi]}}{\sin^2\vartheta} & o & 0\\
-E_{[t\vartheta]} & -E_{[r\vartheta]} & 0 & 0
\end{pmatrix}\,,
\end{equation}
This now further implies that the components \(E_{[t\vartheta]}\), \(E_{[r\vartheta]}\), \(E_{[t\varphi]}\) and \(E_{[r\varphi]}\) vanish identically. After imposing these conditions we further calculate
\begin{equation}
(\cos\varphi\mathcal{L}_{X_y}E - \sin\varphi\mathcal{L}_{X_x}E)_{[\mu\nu]} = \begin{pmatrix}
0 & -\partial_{\vartheta}E_{[tr]} & 0 & 0\\
\partial_{\vartheta}E_{[tr]} & 0 & 0 & 0\\
0 & 0 & 0 & \frac{E_{[\vartheta\varphi]}}{\tan\vartheta} - \partial_{\vartheta}E_{[\vartheta\varphi]}\\
0 & 0 & -\frac{E_{[\vartheta\varphi]}}{\tan\vartheta} + \partial_{\vartheta}E_{[\vartheta\varphi]} & 0
\end{pmatrix}\,,
\end{equation}
from which we deduce that the only remaining components must be of the form
\begin{equation}\label{eq:Espherical}
E_{[tr]} = \tilde{E}_{[tr]}(t, r)\,, \quad
E_{[\vartheta\varphi]} = \tilde{E}_{[\vartheta\varphi]}(t, r)\sin\vartheta\,.
\end{equation}
We finally apply \(X_3\), see \eqref{eq:genvectrans}. Here it is sufficient to calculate
\begin{equation}
(\mathcal{L}_{X_3}E)_{[t\vartheta]} = -\chi\sin\vartheta\tilde{E}_{[tr]}\,, \quad
(\mathcal{L}_{X_3}E)_{[r\varphi]} = \frac{\sin^2\vartheta\tilde{E}_{[\vartheta\varphi]}}{\chi r^2}\,.
\end{equation}
This now shows that also \(E_{[tr]}\) and \(E_{[\vartheta\varphi]}\) vanish, so that the complete antisymmetric part \(E_{[\mu\nu]}\) vanishes. Hence, if we calculate the antisymmetric part of the Euler-Lagrange equations from a tetrad and spin connection with cosmological symmetry, then they are always satisfied, independent of the theory we consider.
\end{proof}

\subsubsection{$\mathrm{ISO}(3)$: flat space $k = 0$}\label{sssec:cosmoflat}
The first case of a cosmological symmetry we discuss here is the spatially flat case \(k = 0\). In this case it is convenient to make use of Cartesian coordinates \(x = r\sin\vartheta\cos\varphi, y = r\sin\vartheta\sin\varphi, z = r\cos\vartheta\), in which the translation generators~\eqref{eq:genvectrans} take the simple form
\begin{equation}\label{eq:genvectransc}
X_1 = \partial_x\,, \quad
X_2 = \partial_y\,, \quad
X_3 = \partial_z\,.
\end{equation}
The elements of the symmetry group are parametrized by \(v \in \mathbb{R}^3\) and \(U \in \mathrm{SO}(3)\), and act on the coordinates as
\begin{equation}
t' = t\,, \quad
\begin{pmatrix}
x'\\
y'\\
z'\\
\end{pmatrix} = U \cdot \begin{pmatrix}
x\\
y\\
z\\
\end{pmatrix} + v\,.
\end{equation}
One easily recognizes the Euclidean group \(\mathrm{ISO}(3) = \mathbb{R}^3 \rtimes \mathrm{SO}(3)\), where the multiplication law of the semidirect product reads \((v, U) \cdot (v', U') = (v + Uv', UU')\).

In order to determine the symmetric tetrads and spin connections, we need to find a homomorphism \(\boldsymbol{\Lambda}: \mathrm{ISO}(3) \to \mathrm{SO}(1,3)\). A natural choice is to enhance the homomorphism~\eqref{eq:rotgrouphomo} to
\begin{equation}\label{eq:eucgrouphomo}
\boldsymbol{\Lambda}: \mathrm{ISO}(3) \to \mathrm{SO}(1,3), (v, U) \mapsto \begin{pmatrix}
1 & 0\\
0 & U\\
\end{pmatrix}
\end{equation}
by simply mapping pure translations \((v, \mathds{1}_3)\) to the identity \(\mathds{1}_4\). The derived Lie algebra homomorphism \(\boldsymbol{\lambda}: \mathfrak{iso}(3) \to \mathfrak{so}(1,3)\) then enhances the homomorphism defined through~\eqref{eq:axalghomo} and~\eqref{eq:rotalghomo} by mapping the translation generators~\eqref{eq:genvectransc} to the zero element \(\mathds{0}_4\).

We then come to solving the symmetry condition~\eqref{eq:infisymcondwb} in the Weitzenböck gauge. Since we have merely extended the spherical symmetry and the homomorphism chosen in section~\ref{ssec:spherical}, we can start from the general spherically symmetric tetrad~\eqref{eq:sphertetradwb} and impose its symmetry under the generators~\eqref{eq:genvectrans}. It already suffices to consider only one of them, such as \(X_3\) as it yields the most simple equations, since the symmetry under the remaining two generators then follows from the structure of the symmetry algebra. Imposing this symmetry yields the conditions
\begin{equation}
C_2 = C_3 = C_6 = 0\,,
\end{equation}
while the remaining parameter functions must be of the form
\begin{equation}
C_1 = n(t)\,, \quad
C_4 = a(t)\,, \quad
C_5 = a(t)r\,.
\end{equation}
Inserting these solutions one obtains the tetrad
\begin{subequations}\label{eq:cosmoftetradwb}
\begin{align}
\theta^0 &= n(t)\dd t\,,\\
\theta^1 &= a(t)\left[\sin\vartheta\cos\varphi\dd r + r\cos\theta\cos\varphi\dd\vartheta - r\sin\vartheta\sin\varphi\dd\varphi\right]\,,\\
\theta^2 &= a(t)\left[\sin\vartheta\sin\varphi\dd r + r\cos\theta\sin\varphi\dd\vartheta + r\sin\vartheta\cos\varphi\dd\varphi\right]\,,\\
\theta^3 &= a(t)\left[\cos\vartheta\dd r - r\sin\vartheta\dd\vartheta\right]\,.
\end{align}
\end{subequations}
Note that this result may be transformed into a diagonal tetrad by applying the local Lorentz transformation
\begin{equation}\label{eq:diagltcosmoflat}
\Lambda^a{}_b = \begin{pmatrix}
1 & 0 & 0 & 0\\
0 & \sin\vartheta\cos\varphi & \sin\vartheta\sin\varphi & \cos\vartheta\\
0 & \cos\vartheta\cos\varphi & \cos\vartheta\sin\varphi & -\sin\vartheta\\
0 & -\sin\varphi & \cos\varphi & 0
\end{pmatrix}\,,
\end{equation}
which is the same as the local Lorentz transformation~\eqref{eq:spherlt1} we introduced in the case of spherical symmetry. One then obtains the tetrad
\begin{equation}\label{eq:diagtetcosmoflat}
\theta'^0 = n(t)\dd t\,, \quad
\theta'^1 = a(t)\dd r\,, \quad
\theta'^2 = a(t)r\dd\vartheta\,, \quad
\theta'^3 = a(t)r\sin\vartheta\dd\varphi\,.
\end{equation}
In this case one has a non-trivial spin connection. Its non-vanishing components are given by
\begin{equation}\label{eq:spiconcosmoflat}
\omega'^1{}_{2\vartheta} = -\omega'^2{}_{1\vartheta} = -1\,, \quad
\omega'^1{}_{3\varphi} = -\omega'^3{}_{1\varphi} = -\sin\vartheta\,, \quad
\omega'^2{}_{3\varphi} = -\omega'^3{}_{2\varphi} = -\cos\vartheta\,,
\end{equation}
which agrees with the spin connection~\eqref{eq:spiconspher1}. Note that this tetrad and spin connection agree with the tetrad~\eqref{eq:tetspher3a} and~\eqref{eq:tetspher3b} as well as the spin connection~\eqref{eq:spiconspher3} with
\begin{equation}
C_t = n(t)\,, \quad
C_r = a(t)\,, \quad
C_s = a(t)r\,, \quad
C_{\upsilon} = C_{\psi} = C_{\alpha} = 0\,.
\end{equation}
Alternatively, one could perform the calculation also in Cartesian coordinates. It then turns out that the obtained tetrad~\eqref{eq:cosmoftetradwb} in the Weitzenböck gauge is simply the diagonal tetrad
\begin{equation}\label{eq:carttetcosmoflat}
\theta^0 = n(t)\dd t\,, \quad
\theta^1 = a(t)\dd x\,, \quad
\theta^2 = a(t)\dd y\,, \quad
\theta^3 = a(t)\dd z\,.
\end{equation}
Note that in these coordinates already the Weitzenböck tetrad is diagonal, so that one has both a diagonal tetrad and a vanishing spin connection.

Also in this case it is useful to calculate a few derived quantities. First note that the determinant of the tetrad is given by
\begin{equation}
\det\theta = n(t)a^3(t)r^2\sin\vartheta\,.
\end{equation}
The metric~\eqref{eq:metric} is, as expected, the flat Robertson-Walker metric
\begin{equation}\label{eq:metcosmoflat}
g_{tt} = -n^2(t)\,, \quad
g_{rr} = a^2(t)\,, \quad
g_{\vartheta\vartheta} = a^2(t)r^2\,, \quad
g_{\varphi\varphi} = g_{\vartheta\vartheta}\sin^2\vartheta\,.
\end{equation}
Finally, the non-vanishing components of the teleparallel affine connection~\eqref{eq:affconn} read
\begin{align}
\Gamma^t{}_{tt} 						&= \frac{n'(t)}{n(t)}\,, &
\Gamma^r{}_{rt} 						&= \Gamma^{\vartheta}{}_{\vartheta t} = \Gamma^{\varphi}{}_{\varphi t} = \frac{a'(t)}{a(t)}\,, &
\Gamma^r{}_{\vartheta\vartheta} 		&= -r\,, &
\Gamma^r{}_{\varphi\varphi} 			&= -r\sin^2\vartheta\,, \nonumber\\
\Gamma^{\vartheta}{}_{r\vartheta} 		&= \Gamma^{\vartheta}{}_{\vartheta r} = \Gamma^{\varphi}{}_{r\varphi} = \Gamma^{\varphi}{}_{\varphi r} = \frac{1}{r}\,, &
\Gamma^{\vartheta}{}_{\varphi\varphi} 	&= -\cos\vartheta\sin\vartheta\,, &
\Gamma^{\varphi}{}_{\vartheta\varphi} 	&= \Gamma^{\varphi}{}_{\varphi\vartheta} = \frac{1}{\tan\vartheta}\,.\label{eq:conncosmoflat}
\end{align}

As we also remarked in the case of spherical symmetry in section~\ref{ssec:spherical}, one still has the freedom to choose a different time coordinate by a coordinate transformation \(t \mapsto \tilde{t}\), since the corresponding diffeomorphisms commute with the cosmological symmetry. This leaves the form of the tetrad~\eqref{eq:diagltcosmoflat} invariant, but the parameter functions transform according to
\begin{equation}\label{eq:cosmocoordtrans}
\tilde{n}(\tilde{t}) = n(t)\frac{\partial t}{\partial\tilde{t}}\,, \quad
\tilde{a}(\tilde{t}) = a(t)\,,
\end{equation}
where \(t\) is now seen as a function of \(\tilde{t}\). One can thus always achieve a constant lapse function \(\tilde{n}(\tilde{t}) = 1\), so that the Robertson-Walker metric~\eqref{eq:metcosmoflat} takes its standard form.

\subsubsection{$\mathrm{SO}(4)$: positively curved space $k = 1$}\label{sssec:cosmopos}
The next case of a cosmological symmetry we study is that of positive spatial curvature \(k = 1\). To understand the geometry of this spacetime, it is helpful to embed the spatial hypersurfaces with constant time \(t\) into \(\mathbb{R}^4\) with coordinates
\begin{equation}
q^1 = r\sin\vartheta\cos\varphi\,, \quad
q^2 = r\sin\vartheta\sin\varphi\,, \quad
q^3 = r\cos\vartheta\,, \quad
q^4 = \sqrt{1 - r^2}\,.
\end{equation}
Note that the spherical coordinates \((t, r, \vartheta, \varphi)\) cover only the positive half space \(q^4 > 0\). In the following we will consider the full space with coordinates \((t, q^A)\), where \(A = 1, \ldots, 4\) and \(q^Aq^B\delta_{AB} = 1\) with \(\delta_{AB} = \mathrm{diag}(1,1,1,1)\), which is thus diffeomorphic to \(\mathbb{R} \times S^3\). The group of spacetime symmetries can then be understood as linear transformations of the spatial coordinates \(q^A\) that leave the Euclidean metric \(\delta_{AB}\) invariant. Hence, they can be parametrized by elements \(U \in \mathrm{SO}(4)\) acting on the spatial coordinates as \(q'^A = U^A{}_Bq^B\).

To construct a homomorphism \(\boldsymbol{\Lambda}: \mathrm{SO}(4) \to \mathrm{SO}(1,3)\) it is helpful to examine the group structure of \(\mathrm{SO}(4)\) and the geometry of \(S^3\). Recall that \(S^3\) is diffeomorphic to the Lie group \(\mathrm{SU}(2)\), where we can use the coordinates \(q^A\) to express
\begin{equation}
Q = iq^k\sigma_k + q^4\mathds{1}_2 = \begin{pmatrix}
q^4 + iq^3 & q^2 + iq^1\\
-q^2 + iq^1 & q^4 - iq^3
\end{pmatrix} \in \mathrm{SU}(2)\,,
\end{equation}
where \(\sigma_k\) are the Pauli matrices and \(k = 1, \ldots, 3\). A rotation \(U \in \mathrm{SO}(4)\) can be decomposed into a pair \((u_L ,u_R)\) with \(u_L, u_R \in \mathrm{SU}(2)\), where the action on \(S^3\) takes the form
\begin{equation}
q'^A = U^A{}_Bq^B \quad \Leftrightarrow \quad Q' = u_LQu_R^{-1}\,.
\end{equation}
This decomposition is unique up to a sign, \((u_L, u_R) \mapsto (-u_L, -u_R)\). Hence, the group \(\mathrm{SU}(2) \times \mathrm{SU}(2)\) is a double cover of \(\mathrm{SO}(4)\). The elements of the two factors are called the left and right isoclinic rotations. We then use the fact that \(\mathrm{SU}(2)\) is a double cover of \(\mathrm{SO}(3)\), i.e., there exists a homomorphism \(\tilde{\bullet}: \mathrm{SU}(2) \to \mathrm{SO}(3), u \mapsto \tilde{u}\). This homomorphism allows us to construct two different homomorphisms \(\boldsymbol{\Lambda}^{\pm}: \mathrm{SO}(4) \to \mathrm{SO}(1,3)\) given by
\begin{equation}
\boldsymbol{\Lambda}^+(u_L, u_R) = \begin{pmatrix}
1 & 0\\
0 & \tilde{u}_L
\end{pmatrix}
\quad \text{and} \quad
\boldsymbol{\Lambda}^-(u_L, u_R) = \begin{pmatrix}
1 & 0\\
0 & \tilde{u}_R
\end{pmatrix}\,.
\end{equation}
These are well-defined, since \(\tilde{u} = \widetilde{-u}\), so that the result does not depend on the choice of the pair \((u_L, u_R)\) or \((-u_L, -u_R)\) to represent \(U \in \mathrm{SO}(4)\). We also remark that pure rotations as discussed in section~\ref{ssec:spherical}, which correspond to those \(U \in \mathrm{SO}(4)\) that leave \(q^4\) fixed, are represented by \(u_L = u_R\), so that both homomorphisms \(\boldsymbol{\Lambda}^{\pm}\) restrict to the same homomorphism~\eqref{eq:rotgrouphomo} on pure rotations.

In the next step we study the induced Lie algebra homomorphisms \(\boldsymbol{\lambda}^{\pm}: \mathfrak{so}(4) \to \mathfrak{so}(1,3)\). Also these restrict to the previously discussed homomorphism given by~\eqref{eq:axalghomo} and~\eqref{eq:rotalghomo} for pure rotations, so that we only have to determine their action on the translation generators~\eqref{eq:genvectrans}. One finds that these are given by
\begin{equation}
\boldsymbol{\lambda}^{\pm}(X_1) = \pm\begin{pmatrix}
0 & 0 & 0 & 0\\
0 & 0 & 0 & 0\\
0 & 0 & 0 & 1\\
0 & 0 & -1 & 0
\end{pmatrix}\,, \quad
\boldsymbol{\lambda}^{\pm}(X_2) = \pm\begin{pmatrix}
0 & 0 & 0 & 0\\
0 & 0 & 0 & -1\\
0 & 0 & 0 & 0\\
0 & 1 & 0 & 0
\end{pmatrix}\,, \quad
\boldsymbol{\lambda}^{\pm}(X_3) = \pm\begin{pmatrix}
0 & 0 & 0 & 0\\
0 & 0 & 1 & 0\\
0 & -1 & 0 & 0\\
0 & 0 & 0 & 0
\end{pmatrix}\,.
\end{equation}
This also shows that the generators of the purely left and right isoclinic rotations given by the combinations \(X_x \pm X_1\), \(X_y \pm X_2\) and \(X_z \pm X_3\) are mapped to the zero element \(\mathds{0}_4\) if one applies the homomorphism \(\boldsymbol{\lambda}^{\mp}\) corresponding to the other subgroup.

Using the homomorphisms \(\boldsymbol{\lambda}^{\pm}\) we can now solve the symmetry condition~\eqref{eq:infisymcondwb} and determine the symmetric tetrads. Again we start from the spherically symmetric tetrad~\eqref{eq:sphertetradwb} discussed in section~\ref{ssec:spherical} and impose symmetry under the translation generators~\eqref{eq:genvectrans}. In this case we find that the parameter functions must be of the form
\begin{equation}
C_1 = n(t)\,, \quad
C_4 = \frac{a(t)}{\chi}\,, \quad
C_5 = r\chi a(t)\,, \quad
C_6 = \mp r^2a(t)\,, \quad
C_2 = C_3 = 0\,.
\end{equation}
where we introduced the abbreviation \(\chi = \sqrt{1 - r^2}\). This yields the two tetrads
\begin{subequations}\label{eq:cosmoptetradwb}
\begin{align}
\theta_{\pm}^0 &= n(t)\dd t\,,\\
\theta_{\pm}^1 &= a(t)\left[\frac{\sin\vartheta\cos\varphi}{\chi}\dd r + r(\chi\cos\theta\cos\varphi \pm r\sin\varphi)\dd\vartheta - r\sin\vartheta(\chi\sin\varphi \mp r\cos\vartheta\cos\varphi)\dd\varphi\right]\,,\\
\theta_{\pm}^2 &= a(t)\left[\frac{\sin\vartheta\sin\varphi}{\chi}\dd r + r(\chi\cos\theta\sin\varphi \mp r\cos\varphi)\dd\vartheta + r\sin\vartheta(\chi\cos\varphi \pm r\cos\vartheta\sin\varphi)\dd\varphi\right]\,,\\
\theta_{\pm}^3 &= a(t)\left[\frac{\cos\vartheta}{\chi}\dd r - r\chi\sin\vartheta\dd\vartheta \mp r^2\sin^2\vartheta\dd\varphi\right]\,,
\end{align}
\end{subequations}
The two solutions we obtained have a simple geometric interpretation. Passing to the inverse tetrads, the vector fields \(e^{\pm}_1, e^{\pm}_2, e^{\pm}_3\) on the spatial hypersurfaces simply correspond to the left or right invariant vector fields on \(\mathrm{SU}(2)\), respectively.

As in the previous case there exists a local Lorentz transformation which allows us to transform the result into a diagonal tetrad. In this case this transformation is given by the matrix
\begin{equation}\label{eq:diagltcosmopos}
\Lambda_{\pm}^a{}_b = \begin{pmatrix}
1 & 0 & 0 & 0\\
0 & \sin\vartheta\cos\varphi & \sin\vartheta\sin\varphi & \cos\vartheta\\
0 & \chi\cos\vartheta\cos\varphi \pm r\sin\varphi & \chi\cos\vartheta\sin\varphi \mp r\cos\varphi & -\chi\sin\vartheta\\
0 & \pm r\cos\vartheta\cos\varphi - \chi\sin\varphi & \chi\cos\varphi \pm r\cos\vartheta\sin\varphi & \mp r\sin\vartheta
\end{pmatrix}\,,
\end{equation}
which we can also write as a product \(\Lambda_{\pm}^a{}_b = \tilde{\Lambda}_{\pm}^a{}_c\Lambda^c{}_b\), where \(\Lambda^a{}_b\) is the previously used local Lorentz transformation~\eqref{eq:diagltcosmoflat} and \(\tilde{\Lambda}_{\pm}^a{}_b\) is given by
\begin{equation}
\tilde{\Lambda}_{\pm}^a{}_b = \begin{pmatrix}
1 & 0 & 0 & 0\\
0 & 1 & 0 & 0\\
0 & 0 & \chi & \mp r\\
0 & 0 & \pm r & \chi
\end{pmatrix}\,.
\end{equation}
Applying this transformation leads to the diagonal tetrad
\begin{equation}\label{diagtetcosmopos}
\theta_{\pm}'^0 = n(t)\dd t\,, \quad
\theta_{\pm}'^1 = \frac{a(t)}{\chi}\dd r\,, \quad
\theta_{\pm}'^2 = a(t)r\dd\vartheta\,, \quad
\theta_{\pm}'^3 = a(t)r\sin\vartheta\dd\varphi\,,
\end{equation}
which is identical for both cases $\pm$. In this case the non-vanishing components of the spin connection distinguish between the sign choices $\pm$, and are given by
\begin{align}
\omega_{\pm}'^1{}_{2\vartheta} 	&= -\omega_{\pm}'^2{}_{1\vartheta} = -\chi\,, &
\omega_{\pm}'^1{}_{2\varphi} 	&= -\omega_{\pm}'^2{}_{1\varphi} = \pm r\sin\vartheta\,, &
\omega_{\pm}'^1{}_{3\vartheta} 	&= -\omega_{\pm}'^3{}_{1\vartheta} = \mp r\,, \nonumber\\
\omega_{\pm}'^1{}_{3\varphi} 	&= -\omega_{\pm}'^3{}_{1\varphi} = -\chi\sin\vartheta\,, &
\omega_{\pm}'^2{}_{3r} 			&= -\omega_{\pm}'^3{}_{2r} = \pm\frac{1}{\chi}\,, &
\omega_{\pm}'^2{}_{3\varphi} 	&= -\omega_{\pm}'^3{}_{2\varphi} = -\cos\vartheta\,.\label{eq:spiconcosmopos}
\end{align}
This tetrad and spin connection agree with the tetrad~\eqref{eq:tetspher3a} and~\eqref{eq:tetspher3b} as well as the spin connection~\eqref{eq:spiconspher3} with
\begin{equation}
C_t = n(t)\,, \quad
C_r = \frac{a(t)}{\chi}\,, \quad
C_s = a(t)r\,, \quad
C_{\upsilon} = C_{\psi} = 0\,, \quad
C_{\alpha} = \mp\arcsin r\,.
\end{equation}

As in the previous case, we calculate a few derived quantities. We find that the determinant of the tetrad is given by
\begin{equation}\label{eq:detcosmopos}
\det\theta = \frac{n(t)a^3(t)r^2\sin\vartheta}{\chi}\,.
\end{equation}
The metric~\eqref{eq:metric} is again of the Robertson-Walker form
\begin{equation}\label{eq:metcosmopos}
g_{tt} = -n^2(t)\,, \quad
g_{rr} = \frac{a^2(t)}{\chi^2}\,, \quad
g_{\vartheta\vartheta} = a^2(t)r^2\,, \quad
g_{\varphi\varphi} = g_{\vartheta\vartheta}\sin^2\vartheta\,,
\end{equation}
and is identical for both sign choices \(\pm\) in the tetrad. Finally, the non-vanishing components of the teleparallel connection~\eqref{eq:affconn} read
\begin{align}
\Gamma^t{}_{tt} 						&= \frac{n'(t)}{n(t)}\,, &
\Gamma^r{}_{rt} 						&= \Gamma^{\vartheta}{}_{\vartheta t} = \Gamma^{\varphi}{}_{\varphi t} = \frac{a'(t)}{a(t)}\,, &
\Gamma^{\vartheta}{}_{r\vartheta} 		&= \Gamma^{\vartheta}{}_{\vartheta r} = \frac{1}{r}\,, &
\Gamma^r{}_{\varphi\varphi} 			&= -r\chi^2\sin^2\vartheta\,, \nonumber\\
\Gamma^r{}_{rr} 						&= \frac{r}{\chi^2}\,, &
\Gamma^r{}_{\vartheta\varphi} 			&= -\Gamma^r{}_{\varphi\vartheta} = \pm r^2\chi\sin\vartheta\,, &
\Gamma^{\vartheta}{}_{\varphi r} 		&= -\Gamma^{\vartheta}{}_{r\varphi} = \pm\frac{\sin\vartheta}{\chi}\,, &
\Gamma^{\varphi}{}_{r\vartheta} 		&= -\Gamma^{\varphi}{}_{\vartheta r} = \pm\frac{1}{\chi\sin\vartheta}\,, \nonumber\\
\Gamma^r{}_{\varphi\varphi} 			&= -r\chi^2\sin^2\vartheta\,, &
\Gamma^{\vartheta}{}_{\varphi\varphi} 	&= -\cos\vartheta\sin\vartheta\,, &
\Gamma^{\varphi}{}_{\vartheta\varphi} 	&= \Gamma^{\varphi}{}_{\varphi\vartheta} = \frac{1}{\tan\vartheta}&
\Gamma^{\varphi}{}_{r\varphi} 			&= \Gamma^{\varphi}{}_{\varphi r} = \frac{1}{r}\,.\label{eq:conncosmopos}
\end{align}
Here we obtain different signs, which shows that the tetrads we found are indeed inequivalent in the sense that they correspond to different affine connections. Note that also here we may eliminate the lapse function \(n(t)\) by applying the coordinate transformation~\eqref{eq:cosmocoordtrans}.

\subsubsection{$\mathrm{SO}(1,3)$: negatively curved space $k = -1$}\label{sssec:cosmoneg}
We finally come to the case \(k = -1\) of negative spatial curvature. We can proceed similarly to the previously discussed case \(k = 1\) of positive spatial curvature, and start by embedding the spatial hypersurfaces into \(\mathbb{R}^4\) with coordinates
\begin{equation}
q^0 = \sqrt{1 + r^2}\,, \quad
q^1 = r\sin\vartheta\cos\varphi\,, \quad
q^2 = r\sin\vartheta\sin\varphi\,, \quad
q^3 = r\cos\vartheta\,.
\end{equation}
The spatial hypersurfaces are then given by the hyperbolic spaces \(\eta_{AB}q^Aq^B = -(q^0)^2 + \delta_{kl}q^kq^l = -1\) with \(q^0 > 0\), where \(A = 0, \ldots, 3\) and \(k, l = 1, \ldots, 3\), so that we use coordinates \((t, q^A)\) for the embedding of the whole spacetime manifold. The group of spacetime symmetries is then given by the linear transformations of the spatial coordinates \(q^A\) that leave the Minkowski metric \(\eta_{AB} = \mathrm{diag}(-1,1,1,1)\) invariant. They can thus be parametrized by elements \(U \in \mathrm{SO}(1,3)\) acting on the spatial coordinates as \(q'^A = U^A{}_Bq^B\). Since in this case the symmetry group is identical to the Lorentz group, the most natural choice of the homomorphism \(\boldsymbol{\Lambda}: \mathrm{SO}(1,3) \to \mathrm{SO}(1,3)\) is simply the identity \(\boldsymbol{\Lambda}^+: U \mapsto U\). Another possible choice is the map \(\boldsymbol{\Lambda}^-: U \mapsto \mathfrak{T}U\mathfrak{T}\), where \(\mathfrak{T} = \mathrm{diag}(-1,1,1,1)\) is the time flip operation. Note that for pure rotations, that leave \(q^0\) invariant, both \(\boldsymbol{\Lambda}^{\pm}\) restrict to the homomorphism~\eqref{eq:rotgrouphomo}. For the derived Lie algebra homomorphisms \(\boldsymbol{\lambda}^{\pm}: \mathfrak{so}(1,3) \to \mathfrak{so}(1,3)\) we thus find, in addition to~\eqref{eq:axalghomo} and~\eqref{eq:rotalghomo}, the relations
\begin{equation}
\boldsymbol{\lambda}^{\pm}(X_1) = \pm\begin{pmatrix}
0 & 1 & 0 & 0\\
1 & 0 & 0 & 0\\
0 & 0 & 0 & 0\\
0 & 0 & 0 & 0
\end{pmatrix}\,, \quad
\boldsymbol{\lambda}^{\pm}(X_2) = \pm\begin{pmatrix}
0 & 0 & 1 & 0\\
0 & 0 & 0 & 0\\
1 & 0 & 0 & 0\\
0 & 0 & 0 & 0
\end{pmatrix}\,, \quad
\boldsymbol{\lambda}^{\pm}(X_3) = \pm\begin{pmatrix}
0 & 0 & 0 & 1\\
0 & 0 & 0 & 0\\
0 & 0 & 0 & 0\\
1 & 0 & 0 & 0
\end{pmatrix}\,.
\end{equation}
We then use this homomorphisms \(\boldsymbol{\lambda}^{\pm}\) to solve the symmetry condition~\eqref{eq:infisymcondwb} in the Weitzenböck gauge. Starting from the spherically symmetric tetrad~\eqref{eq:sphertetradwb} and imposing symmetry under the translation generators~\eqref{eq:genvectrans} yields the conditions
\begin{equation}
C_1 = \pm\chi n(t)\,, \quad
C_2 = \pm\frac{r}{\chi}a(t)\,, \quad
C_3 = rn(t)\,, \quad
C_4 = a(t)\,, \quad
C_5 = r\,, \quad
C_6 = 0\,,
\end{equation}
where we now used \(\chi = \sqrt{1 + r^2}\). This yields the tetrad
\begin{subequations}\label{eq:cosmontetradwb}
\begin{align}
\theta_{\pm}^0 &= \pm n(t)\chi\dd t + \pm a(t)\frac{r}{\chi}\dd r\,,\\
\theta_{\pm}^1 &= a(t)\left[\sin\vartheta\cos\varphi\left(\dd r + \frac{n(t)}{a(t)}r\dd t\right) + r\cos\vartheta\cos\varphi\dd\vartheta - r\sin\vartheta\sin\varphi\dd\varphi\right]\,,\\
\theta_{\pm}^2 &= a(t)\left[\sin\vartheta\sin\varphi\left(\dd r + \frac{n(t)}{a(t)}r\dd t\right) + r\cos\vartheta\sin\varphi\dd\vartheta + r\sin\vartheta\cos\varphi\dd\varphi\right]\,,\\
\theta_{\pm}^3 &= a(t)\left[\cos\vartheta\left(\dd r + \frac{n(t)}{a(t)}r\dd t\right) - r\sin\vartheta\dd\vartheta\right]\,,
\end{align}
\end{subequations}
Note that both tetrads differ only by a (not time orientation preserving) global Lorentz transformation \(\theta_-^a = \mathfrak{T}^a{}_b\theta_+^b\). This is due to the fact that \(\boldsymbol{\Lambda}^+\) and \(\boldsymbol{\Lambda}^-\) differ only by conjugation by \(\mathfrak{T}\), as discussed in section~\ref{ssec:loclor}. We can transform our result to a diagonal tetrad by applying the local Lorentz transformation defined by
\begin{equation}\label{eq:diagltcosmoneg}
\Lambda_{\pm}^a{}_b = \begin{pmatrix}
\pm\chi & r\sin\vartheta\cos\varphi & r\sin\vartheta\sin\varphi & r\cos\vartheta\\
\pm r & \chi\sin\vartheta\cos\varphi & \chi\sin\vartheta\sin\varphi & \chi\cos\vartheta\\
0 & \cos\vartheta\cos\varphi & \cos\vartheta\sin\varphi & -\sin\vartheta\\
0 & -\sin\varphi & \cos\varphi & 0
\end{pmatrix}\,.
\end{equation}
which we can similarly to the case \(k = 1\) write as a product \(\Lambda_{\pm}^a{}_b = \tilde{\Lambda}_{\pm}^a{}_c\Lambda^c{}_b\), where \(\Lambda^a{}_b\) is the previously used local Lorentz transformation~\eqref{eq:diagltcosmoflat} and \(\tilde{\Lambda}_{\pm}^a{}_b\) is now given by
\begin{equation}
\tilde{\Lambda}_{\pm}^a{}_b = \begin{pmatrix}
\pm \chi & r & 0 & 0\\
\pm r & \chi & 0 & 0\\
0 & 0 & 1 & 0\\
0 & 0 & 0 & 1
\end{pmatrix}\,.
\end{equation}
Note that in contrast to the cases~\eqref{eq:diagltcosmoflat} and~\eqref{eq:diagltcosmopos} this Lorentz transformation is not a pure rotation, but also involves a boost. This yields in both cases the same diagonal tetrad
\begin{equation}\label{eq:diagtetcosmoneg}
\theta_{\pm}'^0 = n(t)\dd t\,, \quad
\theta_{\pm}'^1 = \frac{a(t)}{\chi}\dd r\,, \quad
\theta_{\pm}'^2 = a(t)r\dd\vartheta\,, \quad
\theta_{\pm}'^3 = a(t)r\sin\vartheta\dd\varphi\,,
\end{equation}
together with a non-trivial spin connection, whose non-vanishing components take the form
\begin{align}
\omega_{\pm}'^0{}_{1r} 			&= \omega_{\pm}'^1{}_{0r} = \frac{1}{\chi}\,, &
\omega_{\pm}'^0{}_{2\vartheta} 	&= \omega_{\pm}'^2{}_{0\vartheta} = r\,, &
\omega_{\pm}'^0{}_{3\varphi} 	&= \omega_{\pm}'^3{}_{0\varphi} = r\sin\vartheta\,, \nonumber\\
\omega_{\pm}'^1{}_{2\vartheta} 	&= -\omega_{\pm}'^2{}_{1\vartheta} = -\chi\,, &
\omega_{\pm}'^1{}_{3\varphi} 	&= -\omega_{\pm}'^3{}_{1\varphi} = -\chi\sin\vartheta\,, &
\omega_{\pm}'^2{}_{3\varphi} 	&= -\omega_{\pm}'^3{}_{2\varphi} = -\cos\vartheta\,. \label{eq:spiconcosmoneg}
\end{align}
This spin connection was also found in~\cite{Hohmann:2018rwf}. We find that \(\omega_+ = \omega_-\), since the spin connection remains unchanged under global Lorentz transformations. Further, we see that the tetrad~\eqref{eq:diagltcosmoneg} and spin connection~\eqref{eq:spiconcosmoneg} agree with the tetrad~\eqref{eq:tetspher3a} and~\eqref{eq:tetspher3b} as well as the spin connection~\eqref{eq:spiconspher3} with
\begin{equation}
C_t = n(t)\,, \quad
C_r = \frac{a(t)}{\chi}\,, \quad
C_s = a(t)r\,, \quad
C_{\upsilon} = C_{\psi} = \arcsinh r\,, \quad
C_{\alpha} = 0\,.
\end{equation}

Once again we show explicitly a few derived quantities. We find that also in this case the determinant of the tetrad is given by
\begin{equation}\label{eq:detcosmoneg}
\det\theta = \frac{n(t)a^3(t)r^2\sin\vartheta}{\chi}\,,
\end{equation}
which looks similar to the result in the previous section, but we keep in mind that here \(\chi = \sqrt{1 + r^2}\). The metric~\eqref{eq:metric} also takes the Robertson-Walker form
\begin{equation}\label{eq:metcosmoneg}
g_{tt} = -n^2(t)\,, \quad
g_{rr} = \frac{a^2(t)}{\chi^2}\,, \quad
g_{\vartheta\vartheta} = a^2(t)r^2\,, \quad
g_{\varphi\varphi} = g_{\vartheta\vartheta}\sin^2\vartheta\,.
\end{equation}
Finally, the non-vanishing components of the teleparallel connection~\eqref{eq:affconn} read
\begin{align}
\Gamma^t{}_{tt} 						&= \frac{n'(t)}{n(t)}\,, &
\Gamma^r{}_{rt} 						&= \Gamma^{\vartheta}{}_{\vartheta t} = \Gamma^{\varphi}{}_{\varphi t} = \frac{a'(t)}{a(t)}\,, &
\Gamma^r{}_{\vartheta\vartheta} 		&= -r\chi^2\,, &
\Gamma^r{}_{\varphi\varphi} 			&= -r\chi^2\sin^2\vartheta\,, \nonumber\\
\Gamma^r{}_{tr} 						&= \Gamma^{\vartheta}{}_{t\vartheta} = \Gamma^{\varphi}{}_{t\varphi} = \frac{n(t)}{a(t)}\,, &
\Gamma^t{}_{rr} 						&= \frac{a(t)}{n(t)\chi^2}\,, &
\Gamma^t{}_{\vartheta\vartheta} 		&= \frac{a(t)r^2}{n(t)}\,, &
\Gamma^t{}_{\varphi\varphi}				&= \frac{a(t)r^2\sin^2\vartheta}{n(t)}\,, \nonumber\\
\Gamma^r{}_{rr} 						&= -\frac{r}{\chi^2}\,, &
\Gamma^{\vartheta}{}_{r\vartheta} 		&= \Gamma^{\vartheta}{}_{\vartheta r} = \Gamma^{\varphi}{}_{r\varphi} = \Gamma^{\varphi}{}_{\varphi r} = \frac{1}{r}\,, &
\Gamma^{\vartheta}{}_{\varphi\varphi} 	&= -\cos\vartheta\sin\vartheta\,, &
\Gamma^{\varphi}{}_{\vartheta\varphi} 	&= \Gamma^{\varphi}{}_{\varphi\vartheta} = \frac{1}{\tan\vartheta}\,.\label{eq:conncosmoneg}
\end{align}
In this case we obtain the same result for both sign choices \(\pm\) in the tetrad, which agrees with the fact that these two tetrads are related by a global Lorentz transformation. Once again we remark that also here we may eliminate the lapse function \(n(t)\) by applying the coordinate transformation~\eqref{eq:cosmocoordtrans}.

\subsubsection{Complex tetrad for $k = -1$}\label{sssec:cosmonegc}
In a recent article, a complex tetrad was presented for the case of cosmological symmetry with negative spatial curvature \(k = -1\)~\cite{Capozziello:2018hly}. We now show that by extending the formalism we present in this article to allow also complex tetrads and spin connections, one may obtain obtain also this tetrad, as well as another tetrad which differs by a sign. This construction starts with a homomorphism \(\boldsymbol{\Lambda}: \mathrm{SO}(1,3) \to \mathrm{SO}(1,3,\mathbb{C}) \cong \mathrm{SO}(4,\mathbb{C})\), which can be obtained as follows. Let \(U \in \mathrm{SO}(1,3)\), so that \(\eta_{AB}U^A{}_CU^B{}_D = \eta_{CD}\), and construct the complex matrices
\begin{equation}
(\hat{U}^{\pm})^k{}_l = \frac{1}{2}\epsilon^{kmn}\epsilon_{lpq}U_m{}^pU_n{}^q \pm i\epsilon_{lmn}U^{0m}U^{kn} \in \mathrm{GL}(3,\mathbb{C})\,.
\end{equation}
Here we used capital Latin indices \(A, B, \ldots = 0, \ldots, 3\) and small Latin letters from the middle of the alphabet \(k, l, \ldots = 1, \ldots, 3\) as for the real homomorphism discussed in section~\ref{sssec:cosmoneg}. One finds that \(\hat{U}^{\pm} \in \mathrm{SO}(3,\mathbb{C})\). In particular, if \(U \in \mathrm{SO}(3)\) is a pure rotation and has no boost components, one has \(\hat{U} = U\). Finally, we define the group homomorphisms
\begin{equation}
\boldsymbol{\Lambda}^{\pm}: \mathrm{SO}(1,3) \to \mathrm{SO}(1,3,\mathbb{C}), U \mapsto \begin{pmatrix}
1 & 0\\
0 & \hat{U}^{\pm}\\
\end{pmatrix}
\end{equation}
For pure rotations this restricts to~\eqref{eq:rotgrouphomo}. It thus follows that the induced Lie algebra homomorphisms \(\boldsymbol{\lambda}^{\pm}\) extend the relations~\eqref{eq:axalghomo} and~\eqref{eq:rotalghomo} by
\begin{equation}
\boldsymbol{\lambda}^{\pm}(X_1) = \pm\begin{pmatrix}
0 & 0 & 0 & 0\\
0 & 0 & 0 & 0\\
0 & 0 & 0 & i\\
0 & 0 & -i & 0
\end{pmatrix}\,, \quad
\boldsymbol{\lambda}^{\pm}(X_2) = \pm\begin{pmatrix}
0 & 0 & 0 & 0\\
0 & 0 & 0 & -i\\
0 & 0 & 0 & 0\\
0 & i & 0 & 0
\end{pmatrix}\,, \quad
\boldsymbol{\lambda}^{\pm}(X_3) = \pm\begin{pmatrix}
0 & 0 & 0 & 0\\
0 & 0 & i & 0\\
0 & -i & 0 & 0\\
0 & 0 & 0 & 0
\end{pmatrix}\,.
\end{equation}
This finally allows us to solve the symmetry condition~\eqref{eq:infisymcondwb} in the Weitzenböck gauge. Symmetry under the translation generators~\eqref{eq:genvectrans} imposes the conditions
\begin{equation}
C_1 = n(t)\,, \quad
C_4 = \frac{a(t)}{\chi}\,, \quad
C_5 = r\chi a(t)\,, \quad
C_6 = \mp ir^2a(t)\,, \quad
C_2 = C_3 = 0\,.
\end{equation}
on the parameter functions appearing in the spherically symmetric tetrad~\eqref{eq:sphertetradwb}. The sign in \(C_6\) corresponds to the choice of the Lie group homomorphism. Also note that \(C_6\) now becomes imaginary. The tetrad then reads
\begin{subequations}\label{eq:complextetrad}
\begin{align}
\theta_{\pm}^0 &= n(t)\dd t\,,\\
\theta_{\pm}^1 &= a(t)\left[\frac{\sin\vartheta\cos\varphi}{\chi}\dd r + r(\chi\cos\theta\cos\varphi \pm ir\sin\varphi)\dd\vartheta - r\sin\vartheta(\chi\sin\varphi \mp ir\cos\vartheta\cos\varphi)\dd\varphi\right]\,,\\
\theta_{\pm}^2 &= a(t)\left[\frac{\sin\vartheta\sin\varphi}{\chi}\dd r + r(\chi\cos\theta\sin\varphi \mp ir\cos\varphi)\dd\vartheta + r\sin\vartheta(\chi\cos\varphi \pm ir\cos\vartheta\sin\varphi)\dd\varphi\right]\,,\\
\theta_{\pm}^3 &= a(t)\left[\frac{\cos\vartheta}{\chi}\dd r - r\chi\sin\vartheta\dd\vartheta \mp ir^2\sin^2\vartheta\dd\varphi\right]\,.
\end{align}
\end{subequations}
In order to obtain a diagonal tetrad, one applies the (now also complex) local Lorentz transformation
\begin{equation}\label{eq:diagltcosmocom}
\Lambda^a{}_b = \begin{pmatrix}
1 & 0 & 0 & 0\\
0 & \sin\vartheta\cos\varphi & \sin\vartheta\sin\varphi & \cos\vartheta\\
0 & \chi\cos\vartheta\cos\varphi \pm ir\sin\varphi & \chi\cos\vartheta\sin\varphi \mp ir\cos\varphi & -\chi\sin\vartheta\\
0 & \pm ir\cos\vartheta\cos\varphi - \chi\sin\varphi & \chi\cos\varphi \pm ir\cos\vartheta\sin\varphi & \mp ir\sin\vartheta
\end{pmatrix}\,,
\end{equation}
which again can be written as a product \(\Lambda_{\pm}^a{}_b = \tilde{\Lambda}_{\pm}^a{}_c\Lambda^c{}_b\), where \(\Lambda^a{}_b\) is the local Lorentz transformation~\eqref{eq:diagltcosmoflat} and \(\tilde{\Lambda}_{\pm}^a{}_b\) in this case takes the form
\begin{equation}
\tilde{\Lambda}_{\pm}^a{}_b = \begin{pmatrix}
1 & 0 & 0 & 0\\
0 & 1 & 0 & 0\\
0 & 0 & \chi & \mp ir\\
0 & 0 & \pm ir & \chi
\end{pmatrix}\,.
\end{equation}
This yields the same tetrad~\eqref{eq:diagtetcosmoneg} as for the previously described real solution. However, one now finds a complex spin connection
\begin{align}
\omega_{\pm}'^1{}_{2\vartheta} 	&= -\omega_{\pm}'^2{}_{1\vartheta} = -\chi\,, &
\omega_{\pm}'^1{}_{2\varphi} 	&= -\omega_{\pm}'^2{}_{1\varphi} = \pm ir\sin\vartheta\,, &
\omega_{\pm}'^1{}_{3\vartheta} 	&= -\omega_{\pm}'^3{}_{1\vartheta} = \mp ir\,, \nonumber\\
\omega_{\pm}'^1{}_{3\varphi} 	&= -\omega_{\pm}'^3{}_{1\varphi} = -\chi\sin\vartheta\,, &
\omega_{\pm}'^2{}_{3r} 			&= -\omega_{\pm}'^3{}_{2r} = \pm\frac{i}{\chi}\,, &
\omega_{\pm}'^2{}_{3\varphi} 	&= -\omega_{\pm}'^3{}_{2\varphi} = -\cos\vartheta\,,\label{eq:complexspin}
\end{align}
which obviously differs from the (real) spin connection~\eqref{eq:spiconcosmoneg}. This result agrees with the tetrad~\eqref{eq:tetspher3a} and~\eqref{eq:tetspher3b} as well as the spin connection~\eqref{eq:spiconspher3} with
\begin{equation}
C_t = n(t)\,, \quad
C_r = \frac{a(t)}{\chi}\,, \quad
C_s = a(t)r\,, \quad
C_{\upsilon} = C_{\psi} = 0\,, \quad
C_{\alpha} = \mp i\arcsinh r\,.
\end{equation}
In this case we find the same determinant~\eqref{eq:detcosmoneg} of the tetrad and the same Robertson-Walker metric~\eqref{eq:metcosmoneg}. However, we obtain a complex affine connection
\begin{align}
\Gamma^t{}_{tt} 						&= \frac{n'(t)}{n(t)}\,, &
\Gamma^r{}_{rt} 						&= \Gamma^{\vartheta}{}_{\vartheta t} = \Gamma^{\varphi}{}_{\varphi t} = \frac{a'(t)}{a(t)}\,, &
\Gamma^{\vartheta}{}_{r\vartheta} 		&= \Gamma^{\vartheta}{}_{\vartheta r} = \Gamma^{\varphi}{}_{r\varphi} = \Gamma^{\varphi}{}_{\varphi r} = \frac{1}{r}\,, &
\Gamma^r{}_{\varphi\varphi} 			&= -r\chi^2\sin^2\vartheta\,, \nonumber\\
\Gamma^r{}_{rr} 						&= \frac{r}{\chi^2}\,, &
\Gamma^r{}_{\vartheta\varphi} 			&= -\Gamma^r{}_{\varphi\vartheta} = \pm ir^2\chi\sin\vartheta\,, &
\Gamma^{\vartheta}{}_{\varphi r} 		&= -\Gamma^{\vartheta}{}_{r\varphi} = \pm\frac{i\sin\vartheta}{\chi}\,, &
\Gamma^{\varphi}{}_{r\vartheta}			&= -\Gamma^{\varphi}{}_{\vartheta r} = \pm\frac{i}{\chi\sin\vartheta}\,, \nonumber\\
\Gamma^r{}_{\vartheta\vartheta} 		&= -r\chi^2\,, &
\Gamma^{\vartheta}{}_{\varphi\varphi} 	&= -\cos\vartheta\sin\vartheta\,, &
\Gamma^{\varphi}{}_{\vartheta\varphi} 	&= \Gamma^{\varphi}{}_{\varphi\vartheta} = \frac{1}{\tan\vartheta}\,,\label{eq:conncosmonegc}
\end{align}
which obviously differs from the real connection~\eqref{eq:conncosmoneg}, and is more reminiscent of the connection~\eqref{eq:conncosmopos} for the case \(k = 1\), except for the appearance of the imaginary unit in those terms whose sign differs between the different tetrads. As in the previously discussed cases, we may eliminate the lapse function \(n(t)\) by applying the coordinate transformation~\eqref{eq:cosmocoordtrans}.

\subsection{$\mathrm{ISO}(1,3)$, $\mathrm{SO}(1,4)$, $\mathrm{SO}(2,3)$: maximally symmetric spacetimes}\label{ssec:maximal}
We finally come to the case of maximally symmetric spacetimes. These can easily be obtained by enhancing the cosmological symmetry shown in the previous section. This will be shown in section~\ref{ssec:maxsymvect}, where we display the symmetry generating vector fields we use here. The remainder of this section is split in two parts. In section~\ref{ssec:minkowski} we show how to derive a symmetric tetrad for Minkowski spacetime, while in section~\ref{ssec:ads} we prove that no such tetrad exists for (anti-)de Sitter spacetime.

\subsubsection{Symmetry generating vector fields}\label{ssec:maxsymvect}
One possibility to obtain the symmetry generators for maximally symmetric spacetimes is to make use of the previously introduced generators~\eqref{eq:genvecz}, \eqref{eq:genvecxy} and~\eqref{eq:genvectrans}. This can be achieved by choosing the additional vector fields
\begin{subequations}\label{eq:genvecmax}
\begin{align}
X_0 &= \chi\partial_t - kr\chi T_k(t)\partial_r\,,\\
X_X &= r\sin\vartheta\cos\varphi\partial_t + \chi^2\sin\vartheta\cos\varphi T_k(t)\partial_r + \frac{\cos\vartheta\cos\varphi T_k(t)}{r}\partial_{\vartheta} - \frac{\sin\varphi T_k(t)}{r\sin\vartheta}\partial_{\varphi}\,,\\
X_Y &= r\sin\vartheta\sin\varphi\partial_t + \chi^2\sin\vartheta\sin\varphi T_k(t)\partial_r + \frac{\cos\vartheta\sin\varphi T_k(t)}{r}\partial_{\vartheta} + \frac{\sin\varphi T_k(t)}{r\cos\vartheta}\partial_{\varphi}\,,\\
X_Z &= r\cos\vartheta\partial_t + \chi^2\cos\vartheta T_k(t)\partial_r - \frac{\sin\vartheta T_k(t)}{r}\partial_{\vartheta}\,.
\end{align}
\end{subequations}
where \(\chi = \sqrt{1 - kr^2}\) as before and we introduced the shorthand notation
\begin{equation}
T_k(t) = \begin{cases}
\tanh t & \text{for } k = 1\,,\\
t & \text{for } k = 0\,,\\
\tan t & \text{for } k = -1\,.
\end{cases}
\end{equation}
The further treatment depends on the particular choice of \(k\), and hence of the symmetry group of the maximally symmetric spacetime. We will therefore discuss the cases \(k = 0\) and \(k \neq 0\) separately.

\subsubsection{Minkowski spacetime}\label{ssec:minkowski}
The most simple case is given by Minkowski spacetime \(k = 0\), whose symmetry group is \(\mathrm{ISO}(1,3)\). This is a straightforward extension to the case of spatially flat cosmological symmetry with group \(\mathrm{ISO}(3)\) as discussed in section~\ref{sssec:cosmoflat}, so that we can proceed in full analogy. We parametrize the elements of \(\mathrm{ISO}(1,3)\) by \(v \in \mathbb{R}^4\) and \(U \in \mathrm{SO}(1,3)\), so that the action in Cartesian coordinates takes the form
\begin{equation}
\begin{pmatrix}
t'\\
x'\\
y'\\
z'\\
\end{pmatrix} = U \cdot \begin{pmatrix}
t\\
x\\
y\\
z\\
\end{pmatrix} + v\,.
\end{equation}
We then proceed by choosing a homomorphism \(\boldsymbol{\Lambda}: \mathrm{ISO}(1,3) \to \mathrm{SO}(1,3)\). In this case we can simply enhance the homomorphism~\eqref{eq:eucgrouphomo} to
\begin{equation}
\boldsymbol{\Lambda}: \mathrm{ISO}(1,3) \to \mathrm{SO}(1,3), (v, U) \mapsto U\,.
\end{equation}
Note that if \(U\) is a pure rotation (no boosts involved) and \(v^0 = 0\) this reduces to the case discussed in section~\ref{sssec:cosmoflat}. We can therefore make use of the tetrad~\eqref{eq:cosmoftetradwb} and impose in addition symmetry under the generators~\eqref{eq:genvecmax} with \(k = 0\). This yields the solution that \(a(t) = n(t) = c\) with an arbitrary constant \(c\). Also in this case one can apply the Lorentz transformation~\eqref{eq:diagltcosmoflat} in order to obtain the diagonal tetrad~\eqref{eq:diagtetcosmoflat}, together with the non-vanishing spin connection~\eqref{eq:spiconcosmoflat}.

\subsubsection{Anti-de Sitter and de Sitter spacetimes}\label{ssec:ads}
We finally come to the cases of anti-de Sitter and de Sitter spacetimes, whose symmetry groups are \(\mathrm{SO}(2,3)\) and \(\mathrm{SO}(1,4)\), respectively. These cases are fundamentally different from the previously discussed case of Minkowski spacetime. This can be seen in different ways, as we will show in the remainder of this section, in which we prove the following statement:

\begin{stmt}
There are no teleparallel geometries which are symmetric with respect to the generating vector fields~\eqref{eq:genvecz}, \eqref{eq:genvecxy}, \eqref{eq:genvectrans} and~\eqref{eq:genvecmax} of maximal symmetry with \(k = \pm 1\).
\end{stmt}

First note that there exists no non-trivial group homomorphism \(\boldsymbol{\Lambda}: G \to \mathrm{SO}(1,3)\), where \(G\) is one of the two aforementioned groups. This can most easily be seen at the level of the corresponding Lie algebra homomorphism \(\boldsymbol{\lambda}: \mathfrak{g} \to \mathfrak{so}(1,3)\). Since \(\dim\mathfrak{g} = 10 > 6 = \dim\mathfrak{so}(1,3)\), such a homomorphism must necessarily have a non-trivial kernel \(\ker\boldsymbol{\lambda} \subset \mathfrak{g}\). Recall that the kernel of a Lie algebra homomorphism is an ideal of the Lie algebra. However, since \(\mathfrak{g}\) is a real form of the simple Lie algebra \(\mathrm{so}(5, \mathbb{C})\), its only ideals are \(\{0\}\) and \(\mathfrak{g}\) itself. Hence, \(\ker\boldsymbol{\lambda} = \mathfrak{g}\), and so \(\boldsymbol{\lambda}\) is the trivial homomorphism. Inserting this homomorphism into the symmetry condition~\eqref{eq:infisymcondwb} one finds that the resulting system of equations has no solution.

Another possibility to show that there exist no symmetric tetrads in the sense of our definition is by going back to the derivation of the symmetry condition in section~\ref{sec:symtelegrav}. Recall that we required both the metric~\eqref{eq:metric} and the affine connection~\eqref{eq:affconn} to be invariant under the action of the symmetry group on the spacetime manifold. Solving the condition~\eqref{eq:affsymcond} for the generating vector fields~\eqref{eq:genvecz}, \eqref{eq:genvecxy}, \eqref{eq:genvectrans} and~\eqref{eq:genvecmax} shows that the only solution is the Levi-Civita connection of the maximally symmetric metric
\begin{equation}\label{eq:maxsymmetric}
g = -\dd t \otimes \dd t + C_k^2(t)\left[\frac{\dd r \otimes \dd r}{1 - kr^2} + r^2\left(\dd\vartheta \otimes \dd\vartheta + \sin^2\vartheta\dd\varphi \otimes \dd\varphi\right)\right]\,,
\end{equation}
where we used the abbreviation
\begin{equation}
C_k(t) = \begin{cases}
\cosh t & \text{for } k = 1\,,\\
1 & \text{for } k = 0\,,\\
\cos t & \text{for } k = -1\,.
\end{cases}
\end{equation}
However, for \(k \neq 0\) this connection has a non-vanishing curvature, and so it cannot be of the form~\eqref{eq:affconn} with a flat spin connection \(\omega\). Hence, no symmetric teleparallel geometry exists.

We conclude this section with the remark that of course there exist tetrads which correspond to the maximally symmetric metric~\eqref{eq:maxsymmetric} for the de Sitter or anti-de Sitter cases and which appear as solutions to teleparallel gravity theories, even though the corresponding teleparallel connections~\eqref{eq:affconn} do not satisfy the symmetry condition~\eqref{eq:infisymcondgen} for all generators of maximal symmetry. However, these depend on the choice of the particular gravity theory, and a detailed analysis would exceed the scope of this article.

\section{Conclusion}\label{sec:conclusion}
In this article we have discussed the invariance of teleparallel geometries, which are described in terms of a tetrad and a flat Lorentz spin connection, under the action of a group on the underlying spacetime. To do so we interpreted the teleparallel geometry as an orthogonal Cartan geometry and applied the known symmetry concepts from there.

Demanding that the connection is teleparallel, i.e.\ flat, it turned out that this notion of symmetry implies the invariance of the metric as function of the tetrads, the torsion and their covariant derivatives, and hence the field equations of any teleparallel gravity theory built from these quantities, under the group action. We then derived the most general symmetric teleparallel spacetimes for a number of symmetry groups corresponding to
the following symmetries:
\begin{enumerate}
	\item axial symmetry in Weitzenböck gauge \eqref{eq:axtetradwb} and non-Weitzenböck gauge \eqref{eq:axtetradconst},
	\item spherical symmetry in Weitzenböck gauge \eqref{eq:sphertetradwb} and non-Weitzenböck gauge \eqref{eq:tetspher3a} or \eqref{eq:tetspher3b} with spin connection~\eqref{eq:spiconspher3},
	\item homogeneous and isotropic cosmological symmetry with spatial curvature parameters
	\begin{enumerate}
		\item $k=0$:  in Weitzenböck gauge, in spherical \eqref{eq:cosmoftetradwb} and Cartesian coordinates \eqref{eq:carttetcosmoflat}, and as diagonal tetrad \eqref{eq:diagtetcosmoflat} with non-trivial spin connection \eqref{eq:spiconcosmoflat}
		\item $k=1$:  in Weitzenböck gauge \eqref{eq:cosmoptetradwb} and as diagonal tetrad \eqref{diagtetcosmopos} with non-trivial spin connection \eqref{eq:spiconcosmopos}
		\item $k=-1$ real:  in Weitzenböck gauge \eqref{eq:cosmontetradwb} and as diagonal tetrad \eqref{eq:diagtetcosmoneg} with non-trivial spin connection \eqref{eq:spiconcosmoneg}
		\item $k=-1$ complex:  in Weitzenböck gauge \eqref{eq:complextetrad} and as diagonal tetrad \eqref{eq:diagtetcosmoneg} with non-trivial spin connection~\eqref{eq:complexspin}.
	\end{enumerate}
\end{enumerate}
We also displayed the metrics and teleparallel affine connections which are derived from these tetrads and spin connections. Moreover, we briefly discussed how the field equations generically simplify in symmetric situations.

We found that if a teleparallel geometry possesses cosmological symmetry, then it identically solves the antisymmetric part of the field equations of any teleparallel gravity theory. Further, it turned out that for de Sitter and anti-de Sitter spacetimes no symmetric teleparallel geometries exist, since there are no flat affine connections obeying this symmetry. This means that even though one can find tetrads for the maximally symmetric (anti-)de Sitter metric, which may also solve the field equations of particular teleparallel gravity theories, the connection and torsion tensor do not obey the full symmetry group.

The tetrads and spin connections we presented in this article, some of which have previously been obtained attempting to solve (the antisymmetric part of) the field equations of particular teleparallel gravity theories, provide a possible starting point to solve the field equations of any generic teleparallel gravity theory. Our approach has the advantage that it is universal and fully independent of the choice of any particular theory, and it may also be applied to other symmetry groups which are not discussed in this article, such as that of plane waves.

Our results raise a number of interesting questions for future investigation. One such question comes from our finding that for a given symmetry there may in general be different branches of solutions for the tetrad and spin connection, which lead to the same metric, but have different torsion. One would therefore expect that these different solutions can be distinguished only if there is a coupling of matter to torsion, which is not the case for non-spinning test particles. This becomes even more interesting in theories in which the dynamics of the metrics depends on the torsion, and thus differs between these branches, as it is the case for the real and complex tetrad of \(k = -1\) cosmology~\cite{Hohmann:2018rwf,Capozziello:2018hly}, which may hint towards additional degrees of freedom besides the usual metric ones~\cite{Li:2011rn,Nester:2017wau,Ferraro:2018tpu,Hohmann:2018jso,Golovnev:2018wbh,Ferraro:2018axk,Blixt:2018znp}. Further studies are needed in order to understand the physical consequences of these issues.

\begin{acknowledgments}
The authors were supported by the Estonian Research Council through the Institutional Research Funding project IUT02-27 and the Personal Research Funding project PUT790 and PRG356, as well as by the European Regional Development Fund through the Center of Excellence TK133 ``The Dark Side of the Universe''.
\end{acknowledgments}

\bibliographystyle{utphys}
\bibliography{SymmTP}

\end{document}